\documentclass[12pt]{article}
\usepackage{amssymb}
\usepackage{amsmath}
\usepackage{amsfonts}
\usepackage{graphicx}

\input{tcilatex}

\begin{document}

\noindent {\small USC-05/HEP-B2\hfill \hfill hep-th/0512091}\newline
{\small \hfill }

{\vskip0.5cm}

\begin{center}
{\Large Single twistor description of }

{\Large massless, massive, AdS, and other interacting particles}

{\vskip 0.5cm}

\textbf{Itzhak Bars}$^{\dagger }$\textbf{\ and Moises Pic\'{o}n}$^{\dagger
,\ddagger }$

{\vskip 0.4cm}

$^{\dagger }~$\textit{Department of Physics and Astronomy, University of
Southern California, }

\textit{Los Angeles, CA 90089-0484, USA}

$^{\ddagger }$\textbf{~}\textit{Departamento de F\'{\i}sica T\'{e}orica,
Univ. de Valencia and IFIC (CSIC-UVEG),}

\textit{46100-Burjassot (Valencia), Spain}

{\vskip 0.5cm}

\textbf{ABSTRACT}
\end{center}

\noindent The Penrose transform between twistors and the phase space of
massless particles is generalized from the massless case to an assortment of
other particle dynamical systems, including special examples of massless or
massive particles, relativistic or non-relativistic, interacting or
non-interacting, in flat space or curved spaces. Our unified construction
involves always the \textit{same} twistor $Z^{A}$ with only four complex
degrees of freedom and subject to the \textit{same} helicity constraint.
Only the twistor to phase space transform differs from one case to another.
Hence a unification of diverse particle dynamical systems is displayed by
the fact that they all share the same twistor description. Our single
twistor approach seems to be rather different and strikingly economical
construction of twistors compared to other past approaches that introduced
multiple twistors to represent some similar but far more limited set of
particle phase space systems.

\section{Introduction and concepts}

Penrose introduced the twistor program as an alternative to phase space to
give a description of relativistic massless particles in four dimensions
\cite{penrose}\cite{penrose2}. The twistor description is not only Lorentz
covariant, but is also SU$\left( 2,2\right) $ covariant and this makes
evident the well known hidden conformal symmetry SO$\left( 4,2\right) =$SU$%
\left( 2,2\right) $ of the massless system. Penrose's twistors have turned
out to be extremely useful to gain new insight in twistor string theory \cite%
{witten}-\cite{2tstringtwistors} and simplify practical computations in
super Yang-Mills theory \cite{witten2}\cite{cachazo}. We are motivated by
this success to develop a twistor description of more general particle
dynamics which may play a similar useful role in some cases.

In this paper we will use a simple and unified construction of twistors that
will generalize the Penrose transform from the massless particle case to an
assortment of other particle dynamical systems, including special examples
of massless or massive particles, relativistic or non-relativistic,
interacting or non-interacting, in flat space or curved spaces. The basic
formalism for relating phase space and twistor degrees in freedom in any
dimension was given in \cite{2ttwistor} and further developed in \cite%
{2tstringtwistors}\cite{twistorD}. In this paper we apply it explicitly to
examples in four dimensions that have dynamics beyond massless systems. Our
unified construction involves always the \textit{same} four degrees of
freedom in the twistor $Z^{A}=\left( \QATOP{\mu }{\lambda }\right) ,$ $%
A=1,2,3,4,$ constructed from SL$\left( 2,C\right) $ doublet spinors $\mu ^{%
\dot{\alpha}},\lambda _{\alpha },$ each described by two complex degrees of
freedom $\alpha ,\dot{\alpha}=1,2,$ and subject to the \textit{same}
helicity constraint $Z^{A}\bar{Z}_{A}=\mu ^{\dot{\alpha}}\bar{\lambda}_{\dot{%
\alpha}}+\lambda _{\alpha }\bar{\mu}^{\alpha }=2h,$ where $h$ is the
helicity of the particle\footnote{%
At the quantum level the ordered product $\frac{1}{2}(Z^{A}\bar{Z}_{A}+\bar{Z%
}_{A}Z^{A})\psi =2h\psi $ is applied on wavefunctions in twistor space.
Using $\bar{Z}_{A}\psi =-\frac{\partial \psi }{\partial Z^{A}}\psi ,$ this
produces Penrose's homogeneity constraints $Z^{A}\frac{\partial }{\partial
Z^{A}}\psi \left( Z\right) =\left( -2h-2\right) \psi \left( Z\right) $.\label%
{helicityh}}. Only the twistor to phase space transform, which generalizes
the Penrose transform, is different in each one of our examples. This should
be striking in view of previous attempts over many years to construct
twistors for massive particles and other cases that introduced a doubling of
the twistor degrees of freedom \cite{penrose2}, \cite{perjes}-\cite{cederwal}%
. Thus, our economical one-twistor construction, that also unifies a variety
of cases, appears to be a breakthrough in the twistor realm.

We will show that the \textit{same} twistor $Z^{A}$ which is known to
describe the phase space of an on-shell massless particle, is also
equivalent to the phase space of all of the following systems

\begin{itemize}
\item the massless relativistic particle in $d=4$ flat Minkowski space,

\item the massive relativistic particle in $d=4$ flat Minkowski space,

\item the particle on AdS$_{4}$,

\item the particle on AdS$_{3}\times $S$^{1}$, AdS$_{2}\times $S$^{2},$

\item the particle on $R\times S^{3},$

\item the nonrelativistic free particle in 3 space dimensions,

\item the nonrelativistic hydrogen atom in 3 space dimensions,

\item and a related family of other particle systems.
\end{itemize}

Sharing the same twistor equivalent naturally implies duality type
relationships among these systems which may be surprising to some readers.
Moreover, just like the twistor version of massless particles makes the
hidden SU$\left( 2,2\right) $ conformal symmetry a linearly realized evident
symmetry, the twistor versions of all the other particle systems listed
above also make it evident that there is the \textit{same} hidden SU$(2,2)$
symmetry in each one of them. The interpretation of SU$\left( 2,2\right) $
is not conformal transformations of phase space except for the massless
case, but for all cases SU$(2,2)$=SO$(4,2)$ is a hidden symmetry that is
closely related to an underlying $4+2$ dimensional flat spacetime, and
realized in $3+1$ dimensional phase space in different non-linear ways for
each of the systems listed above. In terms of the twistors, the generators
of SU$(2,2)$ are just quadratic and given by the $4\times 4$ matrix $%
J_{~B}^{A}=Z^{A}\bar{Z}_{B}-\frac{1}{4}\delta _{~B}^{A}~\left( Z^{C}\bar{Z}%
_{C}\right) $ which may be expanded in terms of the SO$\left( 4,2\right) $
gamma matrices $\Gamma ^{MN}$
\begin{equation}
J_{~B}^{A}=Z^{A}\bar{Z}_{B}-\frac{1}{4}\delta _{~B}^{A}~\left( Z^{C}\bar{Z}%
_{C}\right) =\frac{1}{4i}L_{MN}~\left( \Gamma ^{MN}\right) _{~B}^{A}.
\label{LZZ}
\end{equation}%
The $J_{~B}^{A}$ satisfy the SU$\left( 2,2\right) $ Lie algebra when the
twistors satisfy the canonical commutation rules $\left[ Z^{A},\bar{Z}_{B}%
\right] =\delta _{~B}^{A}.$ When the twistors $Z^{A}$ are expressed in terms
of the various phase space degrees of freedom of the systems above, we will
show that the SO$(4,2)$ generators $L_{MN}$ take various non-linear phase
space forms and close correctly to form the SO$(4,2)$ Lie algebra under the
phase space Poisson brackets at the classical level\footnote{%
After quantum ordering nonlinear phase space factors, the $L_{MN}$ satisfy
the correct SO$\left( 4,2\right) $ Lie algebra at the quantum level, as
demonstrated for most of these systems elsewhere \cite{2treviews}\cite%
{2tHandAdS}. In this paper we will only discuss the classical level.}.

In this paper we will discuss mainly the zero helicity case $h=0.$ For
general helicity we use precisely the same twistor $Z^{A}$, but the twistor
transform is slightly different due to the helicity, as discussed in detail
elsewhere \cite{twistorsSpin}.

Our results on the hidden SU$(2,2)$ may seem conceptually surprising
independent of twistors: how can a massive particle have SO$(4,2)$ symmetry?
isn't the symmetry of AdS$_{4}$ given by SO$(3,2)$ rather than SO$(4,2)$?
similarly isn't the symmetry of AdS$_{2}\times $S$^{2}$ given by SO$\left(
1,2\right) \times $SO$\left( 3\right) $ rather than SO(4,2), etc.? The
larger unexpected hidden symmetry SO$\left( 4,2\right) $ is indeed difficult
to notice, but its origin was already naturally explained in the context of
two-time physics (2T-physics) \cite{2treviews}\cite{2tHandAdS}. Indeed our
approach in the current paper uses directly previous results of 2T-physics
on the phase spaces of the 1T systems above. We are led to our twistor
results by relating the twistor gauge of 2T-physics to other gauge choices
in phase space. Therefore from the point of view of 2T-physics, the hidden
symmetries which become the evident SU$(2,2)$ acting on the 4-component
twistor $Z^{A}$ are no surprise: This is the same as the SO$(4,2$) acting
linearly on six dimensional phase space $X^{M},P^{M}$ in 2T-physics.

We will give only a very brief description of the concepts of 2T-physics,
and then use it as a technique to construct the relations between the
various phase spaces and the twistors $Z^{A}$. A particle in 1T-physics,
interacting with various backgrounds in $\left( d-1\right) +1$ dimensions
(e.g. electromagnetism, gravity, high spin fields, any potential, etc.), can
be equivalently described in 2T-physics. The 2T theory is in d+2 dimensions,
but has enough gauge symmetry to compensate for the extra 1+1 dimensions, so
that the physical (gauge invariant) degrees of freedom are equivalent to
those encountered in one-time physics (1T-physics). The general 2T theory
for a particle moving in any background field has been constructed \cite%
{2tbacgrounds}%
\begin{equation}
S=\int d\tau ~\left( \dot{X}^{M}P_{M}-\frac{1}{2}A^{ij}Q_{ij}\left(
X,P\right) \right) ,
\end{equation}%
where the symmetric $A^{ij}$ $,$ $i,j=1,2,$ is the Sp$\left( 2,R\right) $
gauge field, and the three $Q_{ij}\left( X,P\right) $ which depend on
background fields, are required to form an Sp$\left( 2,R\right) $ algebra.
The background fields must satisfy certain conditions to comply with the Sp$%
\left( 2,R\right) $ requirement. An infinite number of solutions to the
requirement can be constructed \cite{2tbacgrounds}. So any 1T particle
worldline theory, with any backgrounds, can be obtained as a gauge fixed
version of some 2T particle worldline theory.

The fundamental gauge symmetry in 2T-physics is Sp(2,R) acting on phase
space. A consequence of this gauge symmetry is that position and momentum
become indistinguishable at any instant, so the symmetry is of fundamental
significance. The transformation of $X^{M},P_{M}$ is generally a nonlinear
map that can be explicitly given in the presence of background fields \cite%
{2tbacgrounds}, but in the absence of backgrounds the transformation reduces
to a linear doublet action of Sp$\left( 2,R\right) $ on $\left(
X^{M},P^{M}\right) $ for each $M$ \cite{2treviews}. The physical phase space
is the subspace that is gauge invariant under Sp$\left( 2,R\right) .$ The
gauge invariant subspace of $d+2$ dimensional phase space $X^{M},P_{M}$ is a
phase space in $\left( d-1\right) +1$ dimensions $x^{\mu },p_{\mu }$.
However there are many possible ways to embed the $\left( d-1\right) +1$
phase space in $d+2$ phase space, and this is done by making Sp(2,R) gauge
choices. In the resulting gauge fixed 1T system, time, Hamiltonian, and in
general curved spacetime, are emergent concepts. The Hamiltonian, and
therefore the dynamics as tracked by the emergent time, may look quite
different in one gauge versus another gauge in terms of the remaining gauge
fixed degrees of freedom. In this way, a unique 2T-physics action gives rise
to many 1T-physics dynamical systems.

So, one of the strikingly surprising aspects of 2T physics is that the $d+2$
dimensional theory has many holographic images in $\left( d-1\right) +1$
dimensions. Each image fully captures the gauge invariant physical content
of a unique parent theory, but from the point of view of 1T-physics each
image appears as a different 1T-dynamical system. Thus 2T-physics unifies
many 1T-systems into a family that corresponds to a given 2T-physics parent
in $d+2$ dimensions. The members of such a family naturally must obey
duality-type relationships among them and share many common properties, such
as the same overall global symmetry that may be manifested in hidden
non-linear ways on the fewer $\left( d-1\right) +1$ dimensions.

Thus, 2T-physics can be viewed as a unification approach through higher
dimensions, but distinctly different than Kaluza-Klein theory because there
are no Kaluza-Klein towers of states, but instead there is a family of 1T
systems with duality type relationships among them.

The 1T systems on our list above are members of such a family for $d=4$. In
the present case the parent 2T theory is the simplest version of 2T-physics
without any background fields. The 2T action is \cite{2treviews}
\begin{equation}
S=\frac{1}{2}\int d\tau ~D_{\tau }X_{i}^{M}X_{j}^{N}\eta _{MN}\varepsilon
^{ij}=\int d\tau ~\left( \dot{X}^{M}P^{N}-\frac{1}{2}%
A^{ij}X_{i}^{M}X_{j}^{N}\right) \eta _{MN}.  \label{2Taction}
\end{equation}%
Here $X_{i}^{M}=\left( X^{M}~P^{M}\right) ,$ $i=1,2,$ is a doublet under Sp$%
\left( 2,R\right) $ for every $M,$ the structure $D_{\tau
}X_{i}^{M}=\partial _{\tau }X_{i}^{M}-A_{i}^{~j}X_{j}^{M}$ is the Sp(2,R)
gauge covariant derivative, Sp(2,R) indices are raised and lowered with the
antisymmetric Sp$\left( 2,R\right) $ metric $\varepsilon ^{ij},$ and in the
last expression an irrelevant total derivative $-\left( 1/2\right) \partial
_{\tau }\left( X\cdot P\right) $ is dropped from the action. This action
describes a particle that obeys the Sp(2,R) gauge symmetry, so its momentum
and position are locally indistinguishable due to the gauge symmetry. The Sp$%
\left( 2,R\right) $ constraints $Q_{ij}=X_{i}\cdot X_{j}=0$ that follow from
the equations of motion have non-trivial solutions only if the metric $\eta
_{MN}$ has two timelike dimensions. So for a system in which position and
momentum are locally indistinguishable, to be non-trivial, two timelike
dimensions are necessary as a consequence of the Sp$\left( 2,R\right) $
gauge symmetry.

Thus in our case the target space is flat in $4+2$ dimension, and hence this
system has an SO$(4,2)$ global symmetry. This global symmetry is shared in
the same representation by all the emergent lower dimensional theories in
the same family, and this explains the hidden SO$(4,2)$=SU$(2,2)$ that is
present in all the systems on our list above. Although this is an already
established fact in previous work on 2T-physics, it will come into a new
focus by displaying the explicit twistor$/$phase space transforms given in
this paper. This will be achieved by using the fact that the twistor
description in terms of the 4-component $Z^{A}$ is one of the possible gauge
choices of the same theory \cite{2ttwistor}\cite{2tstringtwistors}\cite%
{twistorD}. From the point of view of twistors this is a striking new
perspective for constructing the twistor equivalent of particle dynamics.
The same method has been applied also in higher dimensions $d=6,10,11$ and
produced twistor equivalents for several interesting spaces, including
supersymmetric AdS$_{5}\times $S$^{5},$ AdS$_{4}\times $S$^{7},$ AdS$%
_{7}\times $S$^{4}$ and others \cite{2tAdSs}\cite{2tsuperstring}\cite%
{2tstringtwistors}.

It is useful to do some simple counting of degrees of freedom. The on shell
phase space $x^{\mu }\left( \tau \right) ,p^{\mu }\left( \tau \right) $ for
a massless particle in four spacetime dimensions, after eliminating one
gauge degree of freedom due to reparametrization of the worldline, and
solving the mass-shell constraint $p^{2}=0,$ has 3 independent positions and
3 independent momentum degrees of freedom. The complex four component
twistor space $Z^{A}$ also has exactly 6 real physical degrees of freedom
after eliminating one overall phase from $Z^{A}$ and solving one real
constraint $Z^{A}\bar{Z}_{A}=2h$. The counting of the phase space physical
degrees of freedom for any particle moving in 3+1 dimensions is also
precisely 6, independent of whether the particle is massless, massive,
moving in flat space or moving in curved space, or interacting with some
potential. Given that the systems listed above all have 6 physical degrees
of freedom, and are already identified in 2T-physics as being holographic
representatives of a unique 4+2 dimensional system, we should expect that
they all must be represented by the same SU(2,2) twistors. This is the idea
that we explore in this paper in order to construct explicitly the relation
between one unique set of twistors $Z^{A}$ and the phase space for each of
the systems listed above.

\section{Twistor space}

The twistor space $Z^{A}=\left( \QATOP{\mu }{\lambda }\right) $ is
classified as the fundamental representation of SU$\left( 2,2\right) ,$ $%
A=1,2,3,4.$ It consists of four complex numbers $\mu ^{\dot{\alpha}},$ $\dot{%
\alpha}=1,2,$ and $\lambda _{\alpha },$ $\alpha =1,2$ which are the two
spinor representations of the Lorentz group SL$\left( 2,C\right) .$ One
defines $\bar{Z}_{A}=\left( \bar{\lambda}_{\dot{\alpha}}~\bar{\mu}^{\alpha
}\right) =Z^{\dagger }\eta ,$ where $\eta =\tau _{1}\times 1$ is the SU$%
\left( 2,2\right) $ metric, and an overbar such as $\bar{\lambda}_{\dot{%
\alpha}}$ means complex conjugation of $\lambda _{\alpha }$. The $Z^{A}$ are
identified up to an overall phase $Z^{A}\sim e^{i\phi }Z^{A}$, and satisfy a
constraint $Z^{A}\bar{Z}_{A}=2h,$ where the constant $h$ is the helicity of
the particle. The irrelevant phase together with the constraint remove two
real degrees of freedom, so that the twistor contains 6 real physical
degrees of freedom. We will show that these are equivalent to the six
physical degrees of freedom of the phase space of a particle in each of the
cases in our list.

In the case of the massless particle the (twistor) $\leftrightarrow $ (phase
space) equivalence is encoded in the Penrose relations (written below for a
helicity zero particle $h=0$ that satisfies $Z^{A}\bar{Z}_{A}=0$ at the
classical level)%
\begin{equation}
\mu ^{\dot{\alpha}}=-ix^{\dot{\alpha}\beta }\lambda _{\beta },\;\lambda
_{\alpha }\bar{\lambda}_{\dot{\beta}}=p_{\alpha \dot{\beta}},
\label{penrose}
\end{equation}%
where the $2\times 2$ Hermitian matrices $x^{\dot{\alpha}\beta },$ $%
p_{\alpha \dot{\beta}}$ are expanded in terms of the Pauli matrices%
\begin{equation}
x^{\dot{\alpha}\beta }\equiv \frac{1}{\sqrt{2}}x^{\mu }\left( \bar{\sigma}%
_{\mu }\right) ^{\dot{\alpha}\beta },\;p_{\alpha \dot{\beta}}\equiv \frac{1}{%
\sqrt{2}}p^{\mu }\left( \sigma _{\mu }\right) _{\alpha \dot{\beta}};\;\sigma
_{\mu }\equiv \left( 1,\vec{\sigma}\right) ,\;\bar{\sigma}_{\mu }\equiv
\left( -1,\vec{\sigma}\right) .
\end{equation}%
So, roughly speaking, $\lambda $ is the square root of momentum $p$, while
position $x$ is the ratio $\mu /\lambda .$ We will give below the analogous
(twistor) $\leftrightarrow $ (phase space) equivalence relations for the
other cases in our list.

The properties of the twistors $Z^{A}$ can be derived from the following $%
\tau $ reparametrization invariant action on the worldline%
\begin{equation}
S=\int d\tau \left( i\bar{Z}_{A}DZ^{A}-2hV\right) ,\;\;DZ^{A}\equiv \frac{%
\partial Z^{A}}{\partial \tau }-iVZ^{A}.  \label{action}
\end{equation}%
Here the 1-form $Vd\tau $ is a U$\left( 1\right) $ gauge field on the
worldline, $DZ^{A}$ is the gauge covariant derivative that satisfies $\delta
_{\varepsilon }\left( DZ^{A}\right) =i\varepsilon \left( DZ^{A}\right) $ for
$\delta _{\varepsilon }V=\partial \varepsilon /\partial \tau $ and $\delta
_{\varepsilon }Z^{A}=i\varepsilon Z^{A}$. Note that the term $2hV$ (absent
in previous literature) is gauge invariant since it transforms as a total
derivative under the infinitesimal gauge transformation$.$ The reason for
requiring the U$\left( 1\right) $ gauge symmetry is the fact that the
overall phase of the $Z^{A}$ is unphysical and drops out in the relation
between phase space and twistors, as in Eq.(\ref{penrose}). Furthermore, the
equation of motion with respect to $V$ imposes the constraint $Z^{A}\bar{Z}%
_{A}-2h=0,$ which is interpreted as the helicity constraint. Taking into
account that $\left( Z^{A}\bar{Z}_{A}-2h\right) $ is the generator of the U$%
\left( 1\right) $ gauge transformations, the meaning of the vanishing
generator (or helicity constraint) is that only the U$\left( 1\right) $
gauge invariant sector of twistor space is physical.

It is a small exercise to show that the twistor defined by this action,
together with the Penrose relation in Eq.(\ref{penrose}) do indeed correctly
describe the phase space of the massless and spinless ($h=0$ at the
classical level) relativistic particle. First, thanks to $\mu ^{\dot{\alpha}%
}=-ix^{\dot{\alpha}\beta }\lambda _{\alpha },$ the constraint is explicitly
satisfied at the classical level%
\begin{equation}
\bar{Z}_{A}Z^{A}=\left( \bar{\lambda}_{\dot{\alpha}}~\bar{\mu}^{\alpha
}\right) \left( \QATOP{\mu ^{\dot{\alpha}}}{\lambda _{\alpha }}\right) =\bar{%
\lambda}_{\dot{\alpha}}\mu ^{\dot{\alpha}}+\bar{\mu}^{\alpha }\lambda
_{\alpha }=-i\bar{\lambda}_{\dot{\alpha}}x^{\dot{\alpha}\beta }\lambda
_{\beta }+i\bar{\lambda}_{\dot{\beta}}x^{\dot{\beta}\alpha }\lambda _{\alpha
}=0.  \label{tw1}
\end{equation}%
Second, the remaining term in the action $S_{0}=i\int d\tau \bar{Z}_{A}\frac{%
\partial Z^{A}}{\partial \tau }=i\int d\tau \left( \bar{\lambda}_{\dot{\alpha%
}}\frac{\partial \mu ^{\dot{\alpha}}}{\partial \tau }+~\bar{\mu}^{\alpha }%
\frac{\partial \lambda _{\alpha }}{\partial \tau }\right) $ that defines the
canonical structure for the twistors $\left[ Z^{A},\bar{Z}_{B}\right]
=\delta _{~B}^{A}$ also correctly defines the canonical structure for the
phase space variables $x^{\mu },p_{\mu }$ by substituting $\mu ^{\dot{\alpha}%
}=-ix^{\dot{\alpha}\beta }\lambda _{\alpha }$ as follows%
\begin{eqnarray}
S_{0} &=&\int d\tau \left[ \bar{\lambda}_{\dot{\alpha}}\frac{\partial }{%
\partial \tau }\left( x^{\dot{\alpha}\beta }\lambda _{\beta }\right) -\bar{%
\lambda}_{\dot{\alpha}}x^{\dot{\alpha}\beta }\frac{\partial \lambda _{\beta }%
}{\partial \tau }\right] =\int d\tau \frac{\partial x^{\dot{\alpha}\beta }}{%
\partial \tau }\lambda _{\beta }\bar{\lambda}_{\dot{\alpha}} \\
&=&\frac{1}{2}\int d\tau \frac{\partial x_{\mu }}{\partial \tau }p_{\nu
}Tr\left( \bar{\sigma}^{\mu }\sigma ^{\nu }\right) =\int d\tau \frac{%
\partial x_{\mu }}{\partial \tau }p_{\mu }.  \label{tw2}
\end{eqnarray}%
This indicates that, aside from factor ordering issues, quantization of
phase space $\left[ x^{\mu },p_{\nu }\right] =i\delta _{~\nu }^{\mu }$ is
consistent with quantization of twistor space $\left[ Z^{A},\bar{Z}_{B}%
\right] =\delta _{~B}^{A}.$ Finally, the form $p_{\alpha \dot{\beta}%
}=\lambda _{\alpha }\bar{\lambda}_{\dot{\beta}}=\frac{1}{\sqrt{2}}p_{\mu
}\left( \sigma ^{\mu }\right) _{\alpha \dot{\beta}}$ automatically satisfies
the mass shell condition $p^{\mu }p_{\mu }=0$ for a massless particle with
positive energy $p_{0}>0,$ as seen explicitly by writing out the matrix form%
\begin{eqnarray}
p_{\alpha \dot{\beta}} &=&\left(
\begin{array}{cc}
\lambda _{1}\lambda _{1}^{\ast } & \lambda _{1}\lambda _{2}^{\ast } \\
\lambda _{2}\lambda _{1}^{\ast } & \lambda _{2}\lambda _{2}^{\ast }%
\end{array}%
\right) =\frac{1}{\sqrt{2}}\left(
\begin{array}{cc}
p^{0}+p_{3} & p_{1}-ip_{2} \\
p+ip_{2} & p^{0}-p_{3}%
\end{array}%
\right) \;  \label{tw31} \\
Tr\frac{p_{\alpha \dot{\beta}}}{\sqrt{2}} &=&\frac{\left\vert \lambda
_{1}\right\vert ^{2}+\left\vert \lambda _{21}\right\vert ^{2}}{\sqrt{2}}%
=p_{0},\;\det \left( p_{\alpha \dot{\beta}}\right) =0=\frac{1}{2}p^{\mu
}p_{\mu }.  \label{tw3}
\end{eqnarray}%
In the remainder of this paper we will show that the same twistor space $%
Z^{A}$ described by the action in Eq.(\ref{action}) satisfies the analogous
(twistor) $\leftrightarrow $ (phase space) equivalence for all the cases on
our list.

\section{General twistor $/$ phase space transform}

In this section we discuss a unified formula that explicitly gives the
substitutes for the Penrose relations of Eq.(\ref{penrose}). The formula was
derived in \cite{2ttwistor}\cite{2tstringtwistors}\cite{twistorD} through
2T-physics techniques. In this paper we simply use it, demonstrate that it
works, and explain the underlying fundamental reasons for its structure.
Thus, the general twistor $/$ phase space transform for \textit{spinless
particles} at the classical level is given by $Z^{A}=\left( \QATOP{\mu ^{%
\dot{\alpha}}}{\lambda _{\alpha }}\right) ,\;$and
\begin{equation}
\mu ^{\dot{\alpha}}=-i\frac{X^{\dot{\alpha}\beta }}{X^{+^{\prime }}}\lambda
_{\beta }\,,\;\;\lambda _{\alpha }\bar{\lambda}_{\dot{\beta}}=\left(
X^{+^{\prime }}P_{\alpha \dot{\beta}}-P^{+^{\prime }}X_{\alpha \dot{\beta}%
}\right) ,  \label{genpenrose}
\end{equation}%
where%
\begin{eqnarray}
X^{\dot{\alpha}\beta } &=&\frac{1}{\sqrt{2}}X^{\mu }\left( \bar{\sigma}_{\mu
}\right) ^{\dot{\alpha}\beta }=\frac{1}{\sqrt{2}}\left(
\begin{array}{cc}
-X^{0}+X^{3} & X_{1}-iX_{2} \\
X_{1}+iX_{2} & -X^{0}-X^{3}%
\end{array}%
\right) ,\;\; \\
P_{\alpha \dot{\beta}} &\equiv &\frac{1}{\sqrt{2}}P^{\mu }\left( \sigma
_{\mu }\right) _{\alpha \dot{\beta}}=\frac{1}{\sqrt{2}}\left(
\begin{array}{cc}
P^{0}+P^{3} & P_{1}-iP_{2} \\
P_{1}+iP_{2} & P^{0}-P^{3}%
\end{array}%
\right) .
\end{eqnarray}%
The $\left( X^{M},P^{M}\right) $ are the SO$\left( 4,2\right) $ vectors of
2T-physics in Eq.(\ref{2Taction}), labelled by $M=\pm ^{\prime },\mu $ or $%
M=0^{\prime },1^{\prime },\mu ,$ and $\mu =\pm ,1,2$ or $\mu =0,1,2,3.$ They
satisfy the fundamental 2T-physics Sp$\left( 2,R\right) $ constraints
\begin{equation}
X\cdot X=P\cdot P=X\cdot P=0.  \label{2Tconstraints}
\end{equation}%
These are the three generators of the gauge symmetry Sp$\left( 2,R\right) $
and their vanishing implies that the physical phase space is the subspace
that is Sp$\left( 2,R\right) $ gauge invariant.

The dot product is constructed with the SO$\left( 4,2\right) $ metric $\eta
^{MN}.$ Evidently the SO$\left( 4,2\right) $ generators $%
L^{MN}=X^{M}P^{N}-X^{N}P^{M}$ commute with the three Sp$\left( 2,R\right) $
generators $X^{2}$, $P^{2}$, $X\cdot P$ since the latter are constructed as
dot products. The metric $\eta ^{MN}$ is conveniently taken in a lightcone
basis $X^{M}=\left( X^{+^{\prime }},X^{-^{\prime }},X^{\mu }\right) $ in the
extra $\left( 1,1\right) $ dimensions$,$ $X^{\pm ^{\prime }}=\frac{1}{\sqrt{2%
}}\left( X^{0^{\prime }}\pm X^{1^{\prime }}\right) ,$ and $\mu =\pm ,1,2$
labels the Minkowski subspace also in a lightcone basis with $X^{\pm }=\frac{%
1}{\sqrt{2}}\left( X^{0}\pm X^{3}\right) $
\begin{eqnarray}
ds^{2} &=&dX^{M}dX^{N}\eta _{MN}=-2dX^{+^{\prime }}dX^{-^{\prime }}+dX^{\mu
}dX^{\nu }\eta _{\mu \nu } \\
&=&-\left( dX^{0^{\prime }}\right) ^{2}+\left( dX^{1^{\prime }}\right)
^{2}-\left( dX^{0}\right) ^{2}+\left( dX^{1}\right) ^{2}+\left(
dX^{2}\right) ^{2}+\left( dX^{3}\right) ^{2} \\
&=&-2dX^{+^{\prime }}dX^{-^{\prime }}-2dX^{+}dX^{-}+\left( dX_{1}\right)
^{2}+\left( dX_{2}\right) ^{2}.  \label{basis}
\end{eqnarray}

The three properties that $Z^{A}=\left( \QATOP{\mu ^{\dot{\alpha}}}{\lambda
_{\alpha }}\right) $ must satisfy are:

\begin{enumerate}
\item First the helicity constraint $\bar{Z}_{A}Z^{A}=0$ for spinless
particles is trivially guaranteed by the phase space / twistor transform in
Eq.(\ref{genpenrose})%
\begin{eqnarray}
\bar{Z}_{A}Z^{A} &=&\left( \bar{\lambda}_{\dot{\alpha}}~\bar{\mu}^{\alpha
}\right) \left( \QATOP{\mu ^{\dot{\alpha}}}{\lambda _{\alpha }}\right) =\bar{%
\lambda}_{\dot{\alpha}}\mu ^{\dot{\alpha}}+\bar{\mu}^{\alpha }\lambda
_{\alpha }  \label{tw11} \\
&=&-i\bar{\lambda}_{\dot{\alpha}}\left( \frac{X^{\dot{\alpha}\beta }}{%
X^{+^{\prime }}}\right) \lambda _{\beta }+i\bar{\lambda}_{\dot{\alpha}%
}\left( \frac{X^{\dot{\alpha}\beta }}{X^{+^{\prime }}}\right) \lambda
_{\beta }=0.
\end{eqnarray}

\item Second, the canonical structure of Eq.(\ref{tw2}) takes the form%
\begin{eqnarray}
S_{0} &=&i\int d\tau \bar{Z}_{A}\partial _{\tau }Z^{A}=i\int d\tau \left[
\bar{\lambda}_{\dot{\alpha}}\partial _{\tau }\mu ^{\dot{\alpha}}+\bar{\mu}%
^{\alpha }\partial _{\tau }\lambda _{\alpha }\right]  \notag \\
&=&\int d\tau \left[ \bar{\lambda}_{\dot{\alpha}}\partial _{\tau }\left(
\frac{X^{\dot{\alpha}\beta }}{X^{+^{\prime }}}\lambda _{\beta }\right) -\bar{%
\lambda}_{\dot{\alpha}}\frac{X^{\dot{\alpha}\beta }}{X^{+^{\prime }}}%
\partial _{\tau }\lambda _{\beta }\right] \\
&=&\int d\tau \partial _{\tau }\left( \frac{X^{\dot{\alpha}\beta }}{%
X^{+^{\prime }}}\right) \lambda _{\beta }\bar{\lambda}_{\dot{\alpha}}=\int
d\tau ~\partial _{\tau }\left( \frac{X^{\mu }}{X^{+^{\prime }}}\right)
\left( X^{+^{\prime }}P_{\mu }-P^{+^{\prime }}X_{\mu }\right)  \label{tw22}
\end{eqnarray}%
and we must show that this form reduces to the canonical structure $S_{0}=$ $%
\int d\tau \left( \dot{x}\cdot p\right) $ for each of the cases in our list.

\item Third, the mass shell condition follows from the form $\lambda
_{\alpha }\bar{\lambda}_{\dot{\beta}}=X^{+^{\prime }}P_{\alpha \dot{\beta}%
}-P^{+^{\prime }}X_{\alpha \dot{\beta}}.$ The left hand side is a rank one
matrix with zero determinant and positive trace as seen from Eq.(\ref{tw31}%
), and this requires that the right hand side must satisfy
\begin{equation}
\left( P^{\mu }-\frac{P^{+^{\prime }}}{X^{+^{\prime }}}X^{\mu }\right)
^{2}=0,\;\left( X^{+^{\prime }}P^{0}-P^{+^{\prime }}X^{0}\right) >0.
\label{tw33}
\end{equation}%
We must show that these imply the mass shell conditions for all the cases in
our list. These properties will be demonstrated in sections (\ref{massive}-%
\ref{other}) below.
\end{enumerate}

It is worth to point out that our twistor formula in Eq.(\ref{genpenrose})
is related to a deeper structure. The expressions in Eq.(\ref{genpenrose})
are equivalent to, and were derived from, the following more insightful Sp$%
\left( 2,R\right) $ gauge invariant expressions \cite{twistorD}
\begin{equation}
\mu ^{\dot{\alpha}}=\left( \QATOP{{L}^{{+}^{\prime }{-}^{\prime }}{+L}^{{+~-}%
}{+iL}^{{12}}}{-\sqrt{{2}}({L}^{{+1}}+i{L}^{{+2}})}\right) \frac{e^{i\phi }}{%
\sqrt{4{L}^{{+}^{\prime }{+}}}},\;\lambda _{\alpha }=\left( \QATOP{2{L}^{{+}%
^{\prime }{+}}}{\sqrt{{2}}({L}^{{+}^{\prime }{1}}+i{L}^{{+}^{\prime }{2}})~}%
\right) \frac{ie^{i\phi }}{\sqrt{4{L}^{{+}^{\prime }{+}}}},  \label{penroseL}
\end{equation}%
Here the $L^{MN}$ are the generators of SO$\left( 4,2\right) $
\begin{equation}
L^{MN}=X^{M}P^{N}-X^{N}P^{M}.
\end{equation}%
We emphasize that $L^{MN}$ are Sp$\left( 2,R\right) $ gauge invariant.
Inserting these $L^{MN}$ into Eq.(\ref{penroseL}), and using the Sp$\left(
2,R\right) $ gauge singlet constraints of Eq.(\ref{2Tconstraints}), gives
the general twistor formula in Eq.(\ref{genpenrose}). Furthermore, we note
that, up to an overall factor, the twistor $Z^{A}=\left( \QATOP{\mu }{%
\lambda }\right) $ in the form of Eq.(\ref{penroseL}) is just the first
column of the 4$\times 4$ matrix
\begin{equation}
\frac{1}{{4}}{L}_{{MN}}{\Gamma }^{{MN}}=\frac{1}{{2}}\left(
\begin{array}{cc}
{L}^{+^{\prime }-^{\prime }}{+}\frac{1}{2}{L}_{\mu \nu }{\bar{\sigma}}^{\mu
\nu } & -i\sqrt{{2}}{L}^{-^{\prime }\mu }~{\bar{\sigma}}_{\mu } \\
i\sqrt{{2}}{L}^{+^{\prime }\mu }~{\sigma }_{\mu } & {-L}^{+^{\prime
}-^{\prime }}{+}\frac{1}{2}{L}_{\mu \nu }{\sigma }^{\mu \nu }%
\end{array}%
\right)   \label{matrixGL}
\end{equation}%
which appears in Eq.(\ref{LZZ}), and is written in an appropriate gamma
matrix basis\footnote{%
The SO$\left( 4,2\right) =$SU$\left( 2,2\right) $ gamma matrices $\Gamma ^{M}
$ in the $4\times 4$ basis, or the $\bar{\Gamma}^{M}$ in the $\bar{4}\times
\bar{4}$ basis, are taken as follows $\Gamma ^{\pm ^{\prime }}=i\sqrt{2}\tau
^{\pm }\times 1$,\ $\Gamma ^{i}=\tau _{3}\times \sigma ^{i}$,\ $\Gamma
^{0}=-1\times 1,$ while $\bar{\Gamma}^{M}$ are the same as the $\Gamma ^{M}$
for $M=\pm ^{\prime },i,$ but for $M=0$ we have $\bar{\Gamma}^{0}=-\Gamma
^{0}=1\times 1.$ With this definition we have $\Gamma ^{M}\bar{\Gamma}%
^{N}+\Gamma ^{N}\bar{\Gamma}^{M}=2\eta ^{MN}$ and $\Gamma ^{MN}=\frac{1}{2}%
\left( \Gamma ^{M}\bar{\Gamma}^{N}-\Gamma ^{N}\bar{\Gamma}^{M}\right) .$
Then $\frac{1}{2}\Gamma _{MN}L^{MN}=-\Gamma ^{+^{\prime }-^{\prime
}}L^{+^{\prime }-^{\prime }}$+~ $\frac{1}{2}L_{\mu \nu }\Gamma ^{\mu \nu }-$
$\Gamma _{~\mu }^{+^{\prime }}L^{-^{\prime }\mu }-$ $\Gamma _{~\mu
}^{-^{\prime }}L^{+^{\prime }\mu }$ takes the form given in Eq.(\ref%
{matrixGL}). This choice of gamma matrices is consistent with the SU$\left(
2,2\right) $ metric $\eta =\tau _{1}\times 1$ used to define $\bar{Z}%
=Z^{\dagger }\eta =\left( \bar{\lambda}~~\bar{\mu}\right) $ for the
fundamental quartet $Z=\left( \QATOP{\mu }{\lambda }\right) .$ According to
this metric we should have $\eta \left( Z\bar{Z}\right) ^{\dagger }\eta =Z%
\bar{Z}.$ Then Eq.(\ref{LZZ}) requires $\eta \left( \Gamma ^{MN}\right)
^{\dagger }\eta =-\Gamma ^{MN},$ and this follows from the property $\eta
\left( \Gamma ^{M}\right) ^{\dagger }\eta =-\bar{\Gamma}^{M}$ satisfied by
our choice of gamma matrices.\label{gamma}} for the $\Gamma ^{M},\bar{\Gamma}%
^{M}$. The other three columns of this matrix give other equivalent forms of
the twistor $Z^{A}$ up to different overall factors. Except for an
undetermined phase, the different overall factors are fixed (as in \ref%
{penroseL}) by requiring that Eq.(\ref{LZZ}) is satisfied. By inserting the
explicit $L^{MN}=X^{[M}P^{N]}$ and using Eq.(\ref{LZZ}) we have
\begin{eqnarray}
\frac{1}{{4i}}{L}_{{MN}}\left( {\Gamma }^{{MN}}\right) _{~B}^{A}~ &=&\frac{1%
}{{2i}}\left[ \left( X\cdot \Gamma \right) \left( P\cdot \bar{\Gamma}\right)
-\left( P\cdot \Gamma \right) \left( X\cdot \bar{\Gamma}\right) \right]
_{~B}^{A}~  \label{LG2} \\
&=&Z^{A}\bar{Z}_{B}-\frac{1}{4}\delta _{~B}^{A}~\left( Z_{C}\bar{Z}%
^{C}\right) .  \label{LG3}
\end{eqnarray}%
From this form one can see that at the classical level $Z$ satisfies the
zero eigenvalue equations $\left( X\cdot \bar{\Gamma}\right) Z=\left( P\cdot
\bar{\Gamma}\right) Z=0$ since $Z$ is proportional to any column of the
matrix (\ref{LG2}) and $X^{M},P^{M}$ are subject to the constraints in Eq.(%
\ref{2Tconstraints}). These zero eigenvalue conditions are equivalent to the
generalized Penrose type transform in Eqs.(\ref{genpenrose},\ref{penroseL}),
and can be further generalized to any dimension \cite{twistorD} directly in
the form of Eq.(\ref{LG2},\ref{LG3}).

Moreover, the structure of these formulas guarantee that the inverse
relation between the $L^{MN}$ and $Z^{A}$ given in Eq.(\ref{LZZ}) also holds
automatically. Hence the formulas in Eq.(\ref{genpenrose}) or Eq.(\ref%
{penroseL}) give the general twistor transform for spinless particles. They
apply not only to massless particles but to all the other cases in our list.
The generalization of these expressions to spinning particles will be
discussed elsewhere.

Now let us comment on some general properties of Sp$(2,R)$ gauge fixing that
will be applied to our formula in order to generate the twistors for the
systems in our list. The $\left( X^{M},P^{M}\right) $ transform as a doublet
under Sp$\left( 2,R\right) $ for every $M\,,$ as can be seen by commuting
the $\left( X^{M},P^{M}\right) $ with Sp$\left( 2,R\right) $ generators in
Eq.(\ref{2Tconstraints})$.$ The $L^{MN}$ are Sp$\left( 2,R\right) $ gauge
invariant since they commute with those generators. Therefore the physical
space consists of arbitrary functions $f\left( L^{MN}\right) $ \cite%
{2treviews}. Since these are all Sp$\left( 2,R\right) $ gauge invariant,
their SO$\left( 4,2\right) $ properties cannot change by making gauge
choices under Sp$\left( 2,R\right) .$ Therefore every holographic picture of
the $\left( 4+2\right) $ phase space that is obtained by making Sp$\left(
2,R\right) $ gauge choices must reproduce the same SO$\left( 4,2\right) $
representation. Hence the physical states and operators can be classified as
traceless tensors $T_{M_{1}\cdots M_{n},N_{1}\cdots N_{n}}\left( L\right) $
of SO$\left( 4,2\right) $ constructed from powers of the antisymmetric $%
L^{MN}.$ These correspond to the double-row Young tableaux $_{\square
\square \square \square }^{\square \square \square \square }\QATOP{\cdots }{%
\cdots }_{\square \square n}^{\square \square n}$ times scalar functions of
the Casimirs\footnote{%
All the SO$\left( d,2\right) $ Casimir eigenvalues in our system $%
L^{MN}=X^{[M}P^{N]}$ vanish in the physical sector at the classical level
due to the constraints in Eq.(\ref{2Tconstraints}). However, they are
non-zero, but fixed numbers, at the quantum level after taking into account
quantum ordering. Thus $C_{2}\left( SO\left( d,2\right) \right) =1-d^{2}/2,$
which corresponds to the singleton representation, as first explained in
\cite{2treviews}\cite{2tfield}. This is an Sp$(2,R)$ gauge invariant result.
Hence, the twistors for all the cases in our list provide another form of
the singleton representation for SO$\left( 4,2\right) $.} of SO$\left(
4,2\right) .$ These tensors can be used, in any Sp$\left( 2,R\right) $ gauge
fixed form to classify the same set of physical states or operators\footnote{%
For a particular application in a 2T-physics gauge useful for
classifying high spin fields, see \cite{vasiliev}.}.

There could also be physical states or operators $f\left( L^{MN}\right) $
that cannot be expanded in powers of $L^{MN}.$ In particular, as we see
above, we can also construct the spinor representation of SO$\left(
4,2\right) $ from the $L^{MN},$ namely the twistor $Z^{A}\left(
L^{MN}\right) $ in Eq.(\ref{penroseL})$.$ This implies that we may construct
physical states from the twistors. Our generalized twistors $Z^{A}$ in Eq.(%
\ref{penroseL}) are functions of the $L^{MN},$ up to the gauge dependent
overall phase $e^{i\phi }$ which could change under the Sp$\left( 2,R\right)
$ gauge transformations. As already noted in the previous section, the
overall phase of the twistor is indeed U$\left( 1\right) $ gauge dependent
and unphysical. However the U$\left( 1\right) $ singlet condition applied on
physical states as explained in footnote (\ref{helicityh}) requires only
homogeneous functions of $Z$, so the physical space is equivalent whether
written in terms of twistors or in terms of the $L^{MN}.$ This key
observation is the underlying reason for the same twistor $Z^{A}$ to be
equivalent to the various phase spaces that are derived by gauge fixing Sp$%
\left( 2,R\right) .$

So, by taking previously obtained solutions of gauge fixing and solving the
constraints of 2T-physics \cite{2treviews}\cite{2tHandAdS}\cite{2tspinning},
and plugging the resulting gauge fixed forms of $\left( X^{M},P^{M}\right) $
into the gauge invariant formula in Eq.(\ref{penroseL}), or its equivalent
in Eq.(\ref{genpenrose}), we will obtain explicit phase space expressions
for our twistor formula. There remains to check that these expressions
explicitly have the properties analogous to those given in Eqs.(\ref{tw1},%
\ref{tw2},\ref{tw3}) as formulated in Eqs.(\ref{tw22},\ref{tw33}). Of
course, these properties are already guaranteed by the symmetries and
structures we have outlined above, but it will be useful and revealing to
demonstrate them explicitly for each case in our list.

Let us first test the general twistor formula for the massless particle
gauge in 2T-physics. In this fixed gauge we have $X^{+^{\prime }}=1$ and $%
P^{+^{\prime }}=0$, and two of the three constraints $X^{2}=X\cdot P=0$ are
solved explicitly as follows
\begin{eqnarray}
X^{M} &=&\left( \overset{+^{\prime }}{1},\;\overset{-^{\prime }}{x^{2}/2}~,\;%
\overset{\mu }{x^{\mu }}\right) ,  \label{massless1} \\
P^{M} &=&\left( ~0~,~x\cdot p~,\;~p^{\mu }\right) .  \label{massless2}
\end{eqnarray}%
The third Sp$\left( 2,R\right) $ gauge choice has not been made in order
keep Lorentz covariance, hence there remains the constraint $%
P^{2}=-2P^{+^{\prime }}P^{-^{\prime }}+P^{\mu }P_{\mu }=p^{2}=0$ to be
imposed, which is the third Sp$\left( 2,R\right) $ generator. Then the
various cross products give the $L^{MN}=X^{M}P^{N}-X^{N}P^{M}$ in the form
\begin{equation}
L^{+^{\prime }-^{\prime }}=x\cdot p,\;L^{+^{\prime }\mu }=p^{\mu },\;L^{\mu
\nu }=x^{\mu }p^{\nu }-x^{\nu }p^{\mu },\;L^{-^{\prime }\mu }=\frac{x^{2}}{2}%
p^{\mu }-x^{\mu }x.p,  \label{conformal2}
\end{equation}%
which are recognized as the generators of SO$\left( 4,2\right) $ conformal
transformations of the $3+1$ dimensional phase space at the classical level.
Inserting these expressions into our general twistor equivalence formula in
Eq.(\ref{penroseL}) or Eq.(\ref{genpenrose}), we derive the Penrose
relations of Eq.(\ref{penrose}), which already satisfy the desired
properties in Eqs.(\ref{tw22},\ref{tw33}).

Furthermore, we can easily check the inverse relation, that the gauge fixed
form of $\mu ,\lambda $ in Eq.(\ref{penrose}) reproduce the $L^{MN}$ through
the general formula of Eq.(\ref{LZZ})%
\begin{eqnarray}
\left( Z\bar{Z}\right) _{~B}^{A} &=&\left(
\begin{array}{cc}
\mu ^{\dot{\alpha}}\bar{\lambda}_{\dot{\beta}} & \mu ^{\dot{\alpha}}\bar{\mu}%
^{\beta } \\
\lambda _{\alpha }\bar{\lambda}_{\dot{\beta}} & \lambda _{\alpha }\bar{\mu}%
^{\beta }%
\end{array}%
\right) =\left(
\begin{array}{cc}
-i\bar{x}p & \bar{x}p\bar{x} \\
p & ip\bar{x}%
\end{array}%
\right)  \label{zzL} \\
&=&\frac{1}{2i}\left(
\begin{array}{cc}
{L}^{+^{\prime }-^{\prime }}{+}\frac{1}{2}{L}^{\mu \nu }{\bar{\sigma}}_{\mu
\nu } & -i\sqrt{{2}}{L}^{-^{\prime }\mu }~{\bar{\sigma}}_{\mu } \\
i\sqrt{{2}}{L}^{+^{\prime }\mu }~{\sigma }_{\mu } & {-L}^{+^{\prime
}-^{\prime }}{+}\frac{1}{2}{L^{\mu \nu }\sigma }_{\mu \nu }%
\end{array}%
\right) .
\end{eqnarray}%
Here the $L^{MN}$ in the last matrix is computed in each block from the
twistors as
\begin{equation}
\mu \bar{\lambda}=-i\bar{x}\lambda \bar{\lambda}=-i\bar{x}p=\frac{1}{2i}%
\left( {L}^{+^{\prime }-^{\prime }}{+}\frac{1}{2}{L}^{\mu \nu }{\bar{\sigma}}%
_{\mu \nu }\right) ,\;\text{etc.,}
\end{equation}%
with ${L}^{+^{\prime }-^{\prime }}=x^{\mu }p_{\mu },$ $L^{\mu \nu }=x^{\mu
}p^{\nu }-x^{\nu }p^{\mu },$ and so on. We see then that the $L^{MN}$
computed from the twistors are precisely the $L^{MN}=X^{M}P^{N}-X^{N}P^{M}$
of Eq.(\ref{conformal2}) computed from the gauge fixed vectors $X^{M},P^{M}$
Eq.(\ref{massless1},\ref{massless2}), in agreement with our general
statement in Eq.(\ref{LZZ}). This test case, combined with the reasoning
provided above, imply that we will succeed just in the same way with the
other 2T-physics gauge choices that correspond to the list in the
introduction. So, each Sp$\left( 2,R\right) $ gauge choice in 2T-physics
will automatically generate the corresponding twistors through our general
formula in Eq.(\ref{genpenrose}), or through the equivalent Sp$\left(
2,R\right) $ gauge invariant formula Eq.(\ref{penroseL}).

\section{Twistors for massive relativistic particles\label{massive}}

We will discuss the massive particle in two versions which correspond to
different looking gauge fixed forms of the $X^{M},P^{M}.$ These are of
course related to each other by Sp$\left( 2,R\right) $ gauge
transformations, but nevertheless seem independently interesting from the
point of view of 1T-physics. An Sp$\left( 2,R\right) $ gauge choice that
describes the massive particle was given in the second paper in \cite%
{2tHandAdS}, and is described in the Appendix. The zero mass limit of this
gauge does not smoothly connect to the massless particle gauge in Eqs.(\ref%
{massless1},\ref{massless2}) and seems to display singularities. This is not
a problem since the connection should be only up to a Sp$\left( 2,R\right) $
gauge transformations. It is possible to perform Sp$\left( 2,R\right) $
gauge transformations to obtain a second form of the massive particle gauge
which has a smooth zero mass limit. We discuss the smooth case in this
section and the non-smooth case in the Appendix.

The smooth massive particle gauge is given by the following gauge choice of
the $X^{M}$, $P^{M}$ components

\begin{eqnarray}
X^{M} &=&\left( \overset{+^{\prime }}{\frac{1+a\mathbf{\;}}{2a\mathbf{\;}}}%
,\;\overset{-^{\prime }}{\;\frac{x^{2}a\mathbf{\;}}{1+a\mathbf{\;}}}~,~~%
\overset{\mu }{x^{\mu }}\right)  \label{massive2x} \\
P^{M} &=&\left( \frac{-m^{2}}{2(x\cdot p)a},\;\;\left( x\cdot p\right) a\;%
\mathbf{,\;\;}p^{\mu }\right)  \label{massive2p}
\end{eqnarray}%
where%
\begin{equation}
a\equiv \sqrt{1+\frac{m^{2}x^{2}}{\left( x\cdot p\right) ^{2}}}.  \label{a}
\end{equation}%
Note that $(x\cdot p)a$ is nonsingular as $x\cdot p\rightarrow 0.$ To arrive
to the above form two gauge choices have been made and the two constraints $%
X\cdot X=X\cdot P$ have been solved, thus eliminating 4 functions. This
fixes completely the four components $X^{\pm ^{\prime }},P^{\pm ^{\prime }}$
in terms of the remaining independent phase space degrees of freedom $x^{\mu
}\left( \tau \right) ,p^{\mu }\left( \tau \right) .$ Note that $x^{\mu
}\left( \tau \right) ,p^{\mu }\left( \tau \right) $ are dynamical variables
while the constant mass $m$ emerges as a \textquotedblleft modulus" from a
gauge fixed version of the other phase space variables $X^{\pm ^{\prime
}},P^{\pm ^{\prime }}$. The third constraints $P\cdot P=0$ gives the mass
shell condition for the massive particle
\begin{equation}
0=P\cdot P=-2P^{+^{\prime }}P^{-^{\prime }}+P^{\mu }P_{\mu }=p^{2}+m^{2}.
\end{equation}%
With this parametrization, the 2T action in Eq.(\ref{2Taction}) reduces to
the action of the relativistic massive particle
\begin{equation}
S=\int d\tau ~\left( \dot{X}^{M}P^{N}-\frac{1}{2}A^{ij}X_{i}^{M}X_{j}^{N}%
\right) \eta _{MN}=\int d\tau \left( \dot{x}^{\mu }p_{\mu }-\frac{1}{2}%
A^{22}\left( p^{2}+m^{2}\right) \right) .
\end{equation}%
and this justifies the parametrization given in Eqs.(\ref{massive2x},\ref%
{massive2p}).

In the limit $m\rightarrow 0,$ as $a\rightarrow 1$ this gauge smoothly
reduces to the massless particle gauge discussed in the previous section. We
should warn the reader that the phase space degrees of freedom $\left(
x^{\mu },p^{\mu }\right) $ of the massive particle are not the same as those
of the massless particle. By applying an Sp$\left( 2,R\right) $ \textit{local%
} transformation, the massive doublets in Eqs.(\ref{massive2x},\ref%
{massive2p}) can be transformed to the massless doublets in Eqs.(\ref%
{massless1},\ref{massless2}). The Sp$\left( 2,R\right) $ gauge
transformation may be regarded as a canonical transformation that includes
the time components (and hence changes the Hamiltonian of the massive
particle to the Hamiltonian of the massless particle).

The SO$(4,2)$ generators $L^{MN}=X^{[M}P^{N]}$ take the following form in
this gauge

\begin{eqnarray}
L^{\mu \nu } &=&x^{\mu }p^{\nu }-x^{\nu }p^{\mu },\text{ \ \ \ }L^{+^{\prime
}-^{\prime }}=\left( x\cdot p\right) a,  \label{LMNmassive1} \\
L^{+^{\prime }\mu } &=&\frac{1+a\mathbf{\;}}{2a\mathbf{\;}}p^{\mu }+\frac{%
m^{2}}{2\left( x\cdot p\right) a}x^{\mu }  \label{LMNmassive2} \\
L^{-^{\prime }\mu } &=&\frac{x^{2}a\mathbf{\;}}{1+a\mathbf{\;}}p^{\mu
}-\left( x\cdot p\right) ax^{\mu }  \label{LMNmassive3}
\end{eqnarray}%
For general $m$ these $L^{MN}$ close under Poisson brackets to form the Lie
algebra of SO$\left( 4,2\right) $ at the classical level. In the massless
limit, $m\rightarrow 0,$ $a\rightarrow 1,$ these $L^{MN}$ reduce to the
familiar expressions for conformal transformations as given in Eq.(\ref%
{conformal2}).

With the appropriate quantum ordering the Lie algebra must close at the
quantum level, and must have the quadratic Casimir eigenvalue for the
singleton representation $C_{2}=1-d^{2}/4,$ which is $C_{2}=-3$ for $d=4,$
to be consistent with SO$\left( 4,2\right) $ covariant quantization of the
system given in \cite{2treviews}.

We study the twistor transform for the massive particle in this gauge by
inserting $X^{+^{\prime }},P^{+^{\prime }},X^{\mu },P^{\mu }$ in Eqs.(\ref%
{massive2x},\ref{massive2p}) into Eq.(\ref{genpenrose}). This gives our new
twistor transform for the massive particle%
\begin{equation}
\mu ^{\dot{\alpha}}=-ix^{\dot{\alpha}\beta }\lambda _{\beta }\frac{2a}{1+a}%
,\;\;\lambda _{\alpha }\bar{\lambda}_{\dot{\beta}}=\frac{1+a\mathbf{\;}}{2a%
\mathbf{\;}}p_{\alpha \dot{\beta}}+\frac{m^{2}}{2(x\cdot p)a}x_{\alpha \dot{%
\beta}}.  \label{twistormassive}
\end{equation}%
We know from Eq.(\ref{tw11}) that this form satisfies automatically the $%
\bar{Z}Z=\bar{\lambda}\mu +\bar{\mu}\lambda =0$ constraint. We now turn to
the canonical structure induced from twistors as formulated in Eq.(\ref{tw22}%
) and compute it in the present gauge%
\begin{eqnarray}
S_{0} &=&i\int d\tau \bar{Z}_{A}\partial _{\tau }Z^{A}=i\int d\tau \left[
\bar{\lambda}_{\dot{\alpha}}\partial _{\tau }\mu ^{\dot{\alpha}}+\bar{\mu}%
^{\alpha }\partial _{\tau }\lambda _{\alpha }\right] \\
&=&\int d\tau \frac{\partial }{\partial \tau }\left( \frac{x^{\mu }}{%
X^{+^{\prime }}}\right) \left( X^{+^{\prime }}p_{\mu }-P^{+^{\prime }}x_{\mu
}\right) \\
&=&\int d\tau \left(
\begin{array}{c}
\left( \partial _{\tau }x\right) \cdot p-\frac{P^{+^{\prime }}}{X^{+^{\prime
}}}\left( \partial _{\tau }x\right) \cdot x \\
-\partial _{\tau }\ln \left( X^{+^{\prime }}\right) \left( x\cdot p-\frac{%
P^{+^{\prime }}}{X^{+^{\prime }}}x\cdot x\right)%
\end{array}%
\right) \\
&=&\int d\tau \left[ \dot{x}\cdot p+\partial _{\tau }\left( \frac{-m^{2}x^{2}%
}{(x\cdot p)\left( 1+a\right) }\right) \right] .
\end{eqnarray}%
The total derivative can be dropped, so $S_{0}=\int d\tau \left( \dot{x}%
\cdot p\right) $ gives the correct canonical structure in phase space. Hence
the canonical twistor space is equivalent to the canonical phase space.

Finally we investigate the mass shell condition that is induced by the
twistors as given in Eq.(\ref{tw33}). Inserting $X^{+^{\prime }}=\frac{1+a%
\mathbf{\;}}{2a\mathbf{\;}}$ and $P^{+^{\prime }}=\frac{-m^{2}}{2(x\cdot p)a}
$ we compute Eq.(\ref{tw33}). In the present case this takes the form%
\begin{eqnarray}
0 &=&\left( p^{\mu }-\frac{P^{+^{\prime }}}{X^{+^{\prime }}}x^{\mu }\right)
^{2}=\left( p^{\mu }+\frac{m^{2}}{(x\cdot p)\left( 1+a\right) }x^{\mu
}\right) ^{2}=\allowbreak \left( p^{2}+m^{2}\right) \\
0 &<&\left( X^{+^{\prime }}p^{0}-P^{+^{\prime }}x^{0}\right) =\frac{1+a}{2a}%
p^{0}+\frac{m^{2}}{2(x\cdot p)a}x^{0}
\end{eqnarray}%
This shows that the twistors in Eq.(\ref{twistormassive}) induce the correct
mass shell condition for a relativistic massive particle. Since $a>0$ the
positivity condition becomes%
\begin{equation}
\frac{m^{2}}{x\cdot p}x^{0}>-\left( 1+a\right) p^{0}.
\end{equation}%
In this equation we write $x\cdot p=-x^{0}p^{0}+\mathbf{x\cdot p}$ and $%
x^{2}=-\left( x^{0}\right) ^{2}+\mathbf{x}^{2},$ insert the on-shell value $%
p^{0}=\pm \sqrt{\mathbf{p}^{2}+m^{2}},$ and then solve for the allowed
region for $x^{0}.$ First consider the limit of large values of $\left\vert
x^{0}\right\vert $ for which $a\left( x^{0}\right) $ has a leading term
independent of $x^{0}$ and is approximated by $a\sim \frac{\left\vert
\mathbf{p}\right\vert }{\sqrt{\mathbf{p}^{2}+m^{2}}}.$ Then $x^{0}$ drops
out and the inequality becomes $\left( m^{2}/p^{0}\right) <(1+\frac{%
\left\vert \mathbf{p}\right\vert }{\sqrt{\mathbf{p}^{2}+m^{2}}})p^{0}.$ This
is satisfied only for positive $p^{0},$ hence the on-shell solution for $%
p^{0}$ is
\begin{equation}
p^{0}>0,\;p^{0}=\sqrt{\mathbf{p}^{2}+m^{2}}.
\end{equation}%
Inserting this into the inequality we search for the allowed regions for $%
x^{0},$ and find that all values of $x^{0}$ are permitted.

Therefore, the twistor representation given in Eq.(\ref{twistormassive})
describes correctly a massive particle of positive energy\footnote{%
If $a$ were taken as the negative square root in Eq.(\ref{a}), then we would
conclude that $p^{0}$ must also be the negative square root.}.

We can now check that the inverse relation also holds, namely that the $%
L^{MN}$ computed from the twistors are precisely the $%
L^{MN}=X^{M}P^{N}-X^{N}P^{M}$ of Eq.(\ref{conformal2}) computed from the
vectors $X^{M},P^{M}.$ Thus, analogous to Eq.(\ref{zzL}) we now compute by
using the twistor transform in Eq.(\ref{twistormassive})%
\begin{eqnarray}
\left( Z\bar{Z}\right) _{~B}^{A} &=&\left(
\begin{array}{cc}
\mu ^{\dot{\alpha}}\bar{\lambda}_{\dot{\beta}} & \mu ^{\dot{\alpha}}\bar{\mu}%
^{\beta } \\
\lambda _{\alpha }\bar{\lambda}_{\dot{\beta}} & \lambda _{\alpha }\bar{\mu}%
^{\beta }%
\end{array}%
\right) \\
&=&\left(
\begin{array}{cc}
-i\bar{x}\left( p+\frac{m^{2}}{(x\cdot p)\left( 1+a\right) }x\right) & \bar{x%
}\left( p+\frac{m^{2}}{(x\cdot p)\left( 1+a\right) }x\right) \bar{x}\frac{2a%
}{1+a} \\
\frac{1+a\mathbf{\;}}{2a\mathbf{\;}}p+\frac{m^{2}}{2(x\cdot p)a}x & i\left(
p+\frac{m^{2}}{(x\cdot p)\left( 1+a\right) }x\right) \bar{x}%
\end{array}%
\right) \\
&=&-i\left(
\begin{array}{cc}
{L}^{+^{\prime }-^{\prime }}{+}\frac{1}{2}{L}^{\mu \nu }{\bar{\sigma}}_{\mu
\nu } & -i\sqrt{{2}}{L}^{-^{\prime }\mu }~{\bar{\sigma}}_{\mu } \\
i\sqrt{{2}}{L}^{+^{\prime }\mu }~{\sigma }_{\mu } & {-L}^{+^{\prime
}-^{\prime }}{+}\frac{1}{2}{L^{\mu \nu }\sigma }_{\mu \nu }%
\end{array}%
\right)
\end{eqnarray}%
where the second line is obtained by manipulations such as
\begin{equation*}
\mu \bar{\lambda}=-i\frac{2a}{1+a}\bar{x}\lambda \bar{\lambda}=-i\frac{2a}{%
1+a}\bar{x}\left( \frac{1+a\mathbf{\;}}{2a\mathbf{\;}}p+\frac{m^{2}}{%
2(x\cdot p)a}x\right) =-i\left( {L}^{+^{\prime }-^{\prime }}{+}\frac{1}{2}{L}%
^{\mu \nu }{\bar{\sigma}}_{\mu \nu }\right) ,\;\text{etc.,}
\end{equation*}%
The $L^{MN}=X^{M}P^{N}-X^{N}P^{M}$ obtained through the twistors in this way
are identical to those given in Eqs.(\ref{LMNmassive1}-\ref{LMNmassive3})
which were computed from the vectors $X^{M},P^{M}$, in agreement with the
general relation in Eq.(\ref{LZZ}).

Hence we have successfully constructed the twistor transform for the massive
particle. We have shown that there is an SO$\left( 4,2\right) =$SU$\left(
2,2\right) $ symmetry that is non-linearly realized in phase space $\left(
x^{\mu },p^{\mu }\right) ,$ but is linearly realized and is an explicit
symmetry of the action in the twistor version Eq.(\ref{action}) or
2T-physics version Eq.(\ref{2Taction}). The SU$\left( 2,2\right) $ symmetry
is identical for the massive or the massless particle, and in both cases
corresponds to the unitary singleton representation of SO$\left( 4,2\right) $
whose Casimir eigenvalues are independent of the mass parameter (e.g. $%
C_{2}=0$ at the classical level, but $C_{2}=-3$ at the quantum level).
However, the same SO$\left( 4,2\right) $ is realized in rather different
forms in the phase spaces of massive versus massless particles. This
unfamiliar result, which was discovered in 2T-physics, has now taken a new
facade through twistor space.

\section{Twistors for nonrelativistic particle}

The nonrelativistic particle is given by the following gauge choice of the $%
X^{M}$, $P^{M}$ components in $\left( 4+2\right) $ dimensional phase space
in 2T-physics (see second paper in \cite{2tHandAdS})

\begin{eqnarray}
X^{M} &=&\overset{+^{\prime }\;\;\;\;-^{\prime
}\;\;\;\;\;\;\;\;\;\;0\;\;i=1,2,3}{\left( t,~\frac{\mathbf{r\cdot p-}tH}{m}%
,\;u~,\;\mathbf{r}^{i}\right) },  \label{nrx} \\
P^{M} &=&\left( \;m,\;\;H\;\;\mathbf{,\;\;\;}0\;,~\mathbf{p}^{i}\right)
\label{nrp}
\end{eqnarray}%
where $u$ is fixed by%
\begin{equation}
\;u^{2}\equiv \mathbf{r}^{2}-\frac{2t}{m}\mathbf{r\cdot p+}\frac{2t^{2}}{m}H.
\label{u}
\end{equation}%
To arrive to the above form two gauge choices have been made: $P^{0}\left(
\tau \right) =0,$ and $P^{+^{\prime }}\left( \tau \right) =m$ a constant,
for all $\tau $. Solving the two constraints $X\cdot X=X\cdot P=0$ fixes
completely $X^{-^{\prime }}$ and $X^{0}$ in terms of the phase space
variables $\left( t\left( \tau \right) ,\mathbf{r}^{i}\left( \tau \right)
\right) $ and $\left( H\left( \tau \right) ,\mathbf{p}^{i}\left( \tau
\right) \right) $ as given above. The third constraint becomes $P\cdot
P=-2mH+\mathbf{p}^{2}=0,$ which implies that $H=\mathbf{p}^{2}/2m$ is the
Hamiltonian for the non-relativistic particle. With this parametrization,
the 2T action in Eq.(\ref{2Taction}) reduces to the action of the
nonrelativistic massive particle
\begin{eqnarray}
S &=&\int d\tau ~\left( \dot{X}^{M}P^{N}-\frac{1}{2}A^{ij}X_{i}^{M}X_{j}^{N}%
\right) \eta _{MN} \\
&=&\int d\tau \left( -\dot{t}H+\mathbf{\dot{r}}\cdot \mathbf{p}-\frac{1}{2}%
A^{22}\left( -2mH+\mathbf{p}^{2}\right) \right) .
\end{eqnarray}%
This justifies the parametrization given in Eq.(\ref{nrx},\ref{nrp}). If the
remaining gauge freedom is fixed as $t\left( \tau \right) =\tau ,$ and the
constraint $0=P\cdot P=-2mH+\mathbf{p}^{2}$ is imposed explicitly, this
action becomes the standard action of the non-relativistic particle
\begin{equation}
S=\int d\tau \left( \mathbf{\dot{r}}\cdot \mathbf{p}-\frac{\mathbf{p}^{2}}{2m%
}\right) .
\end{equation}%
In this case $X^{-^{\prime }},X^{0}$ become $X^{-^{\prime }}=\mathbf{p\cdot
r-\tau }\frac{\mathbf{p}^{2}}{2m}$ and $X^{0}=u=\pm \left\vert \mathbf{%
r-\tau }\frac{\mathbf{p}}{m}\right\vert $ respectively.

Twistors may be discussed either in the fully gauge fixed form or in the
partially gauge fixed form, but it is preferable not to choose the last
gauge and treat $\left( t,H\right) $ as canonical variables, while applying
the constraint $H-\frac{\mathbf{p}^{2}}{2m}=0$ on physical states.

The SO$(4,2)$ generators $L^{MN}=X^{[M}P^{N]}$ are%
\begin{eqnarray}
L^{ij} &=&\mathbf{r}^{i}\mathbf{p}^{j}-\mathbf{r}^{j}\mathbf{p}%
^{i},\;\;L^{+^{\prime }-^{\prime }}=2tH-\mathbf{r\cdot p,\;}L^{+^{\prime
}i}=t\mathbf{p}^{i}-m\mathbf{r}^{i}  \label{LMNnr1} \\
\mathbf{\;}L^{+^{\prime }0} &=&-mu,\;\;L^{-^{\prime }0}=-Hu,\;\;L^{-^{\prime
}i}=\frac{\mathbf{r\cdot p}}{m}\mathbf{p}^{i}-H\left( \mathbf{r}^{i}+\frac{t%
}{m}\mathbf{p}^{i}\right) ,\;  \label{LMNnr2}
\end{eqnarray}%
One can check that by using the Poisson brackets $\left\{ t,H\right\}
=-1,\;\left\{ \mathbf{r}^{i},\mathbf{p}^{j}\right\} =\delta ^{ij}$ the Lie
algebra of SO$\left( 4,2\right) $ is satisfied at the classical level.
Alternatively, in the gauge $t\left( \tau \right) =\tau $ and $H=\frac{%
\mathbf{p}^{2}}{2m}$, we use only $\left\{ \mathbf{r}^{i},\mathbf{p}%
^{j}\right\} =\delta ^{ij}$ and treat $\tau $ as a parameter, to show that
the Lie algebra of SO$\left( 4,2\right) $ is satisfied at the classical
level at any $\tau .$ It is harder to establish the Lie algebra at the
quantum level due to the non-linear ordering problems presented by the
square root in $u=\pm \left( \mathbf{r}^{2}-\frac{2t}{m}\mathbf{r\cdot p+}%
\frac{2t^{2}}{m}H\right) ^{1/2}.$

We construct the new twistor transform for the nonrelativistic particle by
inserting $X^{+^{\prime }},P^{+^{\prime }},X^{\mu },P^{\mu }$ in Eqs.(\ref%
{nrx},\ref{nrp}) into Eq.(\ref{genpenrose}). This gives
\begin{equation}
\mu ^{\dot{\alpha}}=-i\frac{1}{t}\left( -u+\mathbf{\vec{r}\cdot \vec{\sigma}}%
\right) ^{\dot{\alpha}\beta }\lambda _{\beta },\;\;\lambda _{\alpha }\bar{%
\lambda}_{\dot{\beta}}=\left( mu+\left( t\mathbf{\vec{p}\mathbf{-}}m\mathbf{%
\mathbf{\vec{r}}}\right) \mathbf{\cdot \vec{\sigma}}\right) _{\alpha \dot{%
\beta}}.  \label{twistornr}
\end{equation}%
We know from Eq.(\ref{tw11}) that the $\bar{Z}Z=\bar{\lambda}\mu +\bar{\mu}%
\lambda =0$ constraint for spinless particles is automatically satisfied. We
now turn to the canonical structure induced from twistors as formulated in
Eq.(\ref{tw22}) and compute it in the present gauge%
\begin{eqnarray}
S_{0} &=&i\int d\tau \bar{Z}_{A}\partial _{\tau }Z^{A}=i\int d\tau \left[
\bar{\lambda}_{\dot{\alpha}}\partial _{\tau }\mu ^{\dot{\alpha}}+\bar{\mu}%
^{\alpha }\partial _{\tau }\lambda _{\alpha }\right] \\
&=&\int d\tau \frac{\partial }{\partial \tau }\left( \frac{X^{\mu }}{%
X^{+^{\prime }}}\right) \left( X^{+^{\prime }}P_{\mu }-P^{+^{\prime }}X_{\mu
}\right) \\
&=&\int d\tau \left( \partial _{\tau }\left( \frac{u}{t}\right) mu+\partial
_{\tau }\left( \frac{1}{t}\mathbf{\vec{r}}\right) \cdot \left( t\mathbf{\vec{%
p}\mathbf{-}}m\mathbf{\mathbf{\vec{r}}}\right) \right) \\
&=&\int d\tau \left[ -\dot{t}H+\mathbf{\dot{r}}\cdot \mathbf{p}+\partial
_{\tau }\left( \frac{mu^{2}-m\mathbf{r}^{2}}{2t}\right) \right] .
\end{eqnarray}%
The total derivative can be dropped, so $S_{0}$ gives the correct canonical
structure in phase space.

Next we investigate the mass shell condition that is induced by the twistors
as given in Eq.(\ref{tw33}). In the present case this takes the form%
\begin{eqnarray}
0 &=&\left( P^{\mu }-\frac{P^{+^{\prime }}}{X^{+^{\prime }}}X^{\mu }\right)
^{2}=-\left( \frac{m}{t}u\right) ^{2}+\left( \mathbf{p}-\frac{m}{t}\mathbf{r}%
\right) ^{2},\;\; \\
0 &<&\left( X^{+^{\prime }}P^{0}-P^{+^{\prime }}X^{0}\right) =-mu
\end{eqnarray}%
From this we see that we must choose the negative square root for the $u^{2}$
given in Eq.(\ref{u})
\begin{equation}
u=-\left( \mathbf{r}^{2}-\frac{2t}{m}\mathbf{r\cdot p+}\frac{2t^{2}}{m}%
H\right) ^{1/2}
\end{equation}%
Inserting this form we find%
\begin{equation}
0=-\left( \frac{m}{t}u\right) ^{2}+\left( \mathbf{p}-\frac{m}{t}\mathbf{r}%
\right) ^{2}=-2mH+\mathbf{p}^{2}
\end{equation}%
This shows that the correct mass shell condition is induced by the twistors
in Eq.(\ref{twistornr}). After using $H=\mathbf{p}^{2}/2m$ we find $%
u=-\left\vert \mathbf{r-}t\frac{\mathbf{p}}{m}\right\vert ,$ which shows
there is no condition on the range of $t$ for the square root to be real.
Therefore, the twistor representation given in Eq.(\ref{twistornr})
describes correctly a nonrelativistic particle.

Next we investigate the inverse relation analogous to Eq.(\ref{zzL}). Using
the twistor transform in Eq.(\ref{twistornr}) we have%
\begin{equation}
\left( Z\bar{Z}\right) _{~B}^{A}=\left(
\begin{array}{cc}
\mu ^{\dot{\alpha}}\bar{\lambda}_{\dot{\beta}} & \mu ^{\dot{\alpha}}\bar{\mu}%
^{\beta } \\
\lambda _{\alpha }\bar{\lambda}_{\dot{\beta}} & \lambda _{\alpha }\bar{\mu}%
^{\beta }%
\end{array}%
\right) =\frac{1}{2i}\left(
\begin{array}{cc}
{L}^{+^{\prime }-^{\prime }}{+}\frac{1}{2}{L}^{\mu \nu }{\bar{\sigma}}_{\mu
\nu } & -i\sqrt{{2}}{L}^{-^{\prime }\mu }~{\bar{\sigma}}_{\mu } \\
i\sqrt{{2}}{L}^{+^{\prime }\mu }~{\sigma }_{\mu } & {-L}^{+^{\prime
}-^{\prime }}{+}\frac{1}{2}{L^{\mu \nu }\sigma }_{\mu \nu }%
\end{array}%
\right)  \label{ZZLnr}
\end{equation}%
where
\begin{eqnarray}
\mu \bar{\lambda} &=&-i\frac{1}{t}\left( -u+\mathbf{\vec{r}\cdot \vec{\sigma}%
}\right) \left( mu+\left( t\mathbf{\vec{p}\mathbf{-}}m\mathbf{\mathbf{\vec{r}%
}}\right) \mathbf{\cdot \vec{\sigma}}\right) , \\
\mu \bar{\mu} &=&\;\frac{1}{t^{2}}\left( -u+\mathbf{\vec{r}\cdot \vec{\sigma}%
}\right) \left( mu+\left( t\mathbf{\vec{p}\mathbf{-}}m\mathbf{\mathbf{\vec{r}%
}}\right) \mathbf{\cdot \vec{\sigma}}\right) \left( -u+\mathbf{\vec{r}\cdot
\vec{\sigma}}\right) , \\
\lambda \bar{\lambda} &=&\left( mu+\left( t\mathbf{\vec{p}\mathbf{-}}m%
\mathbf{\mathbf{\vec{r}}}\right) \mathbf{\cdot \vec{\sigma}}\right) , \\
\lambda \bar{\mu} &=&i\frac{1}{t}\left( mu+\left( t\mathbf{\vec{p}\mathbf{-}}%
m\mathbf{\mathbf{\vec{r}}}\right) \mathbf{\cdot \vec{\sigma}}\right) \left(
-u+\mathbf{\vec{r}\cdot \vec{\sigma}}\right) .
\end{eqnarray}%
The $L^{MN}=X^{M}P^{N}-X^{N}P^{M}$ obtained through the twistors by
comparing the matrices in Eq.(\ref{ZZLnr}) are identical to those given in
Eqs.(\ref{LMNnr1}-\ref{LMNnr2}) which were computed directly from the
vectors $X^{M},P^{M}$. This is in agreement with the general relation in Eq.(%
\ref{LZZ}).

Hence we have successfully constructed the twistor transform for the massive
particle. This shows that the SU$\left( 2,2\right) $ symmetry, that is
linearly realized on the twistors $Z^{A}$ or on the vectors $\left(
X^{M},P^{M}\right) ,$ is a non-linearly realized hidden symmetry of the
action $S=\int d\tau \left( \mathbf{\dot{r}}\cdot \mathbf{p}-\frac{\mathbf{p}%
^{2}}{2m}\right) $ of the non-relativistic particle, which indeed is the
case (see second paper in \cite{2tHandAdS}). Furthermore, there exists a
duality among the nonrelativistic particle, the relativistic massive or
massless particles as well as all the other particle systems discussed in
this paper, since they are all constructed from the same twistor $Z^{A}$, or
the same 2T phase space $X^{M},P^{M},$ and they all are realizations of the
singleton representation of SO$\left( 4,2\right) .$

\section{Twistors for particles in AdS$_{4-n}\times $S$^{n}$}

We consider a particle moving in AdS$_{d-n}\times $S$^{n}$ spaces for $%
n=0,1,2,\cdots ,\left( d-2\right) $. In general, to describe these spaces we
make the following Sp$\left( 2,R\right) $ gauge choice in $\left( d+2\right)
$ dimensional phase space in 2T-physics (see second paper in \cite{2tHandAdS}%
)

\begin{eqnarray}
X^{M} &=&\frac{R}{\left\vert \mathbf{y}\right\vert }\left( \overset{%
+^{\prime }}{R~~},\;~~\overset{-^{\prime }}{\frac{x^{2}+y^{2}}{2R}},~~\;~~~~%
\overset{m}{x^{m~}}~,~~\overset{I}{\mathbf{y}^{I}}\right) ,\;\;\QATOP{%
{\small m=0,1,\cdots ,}\left( {\small d-2-n}\right) {\small ~~~~~}}{{\small %
I=1,\cdots ,(n+1)~~~~~~~}}  \label{AdSS1} \\
P^{M} &=&\frac{\left\vert \mathbf{y}\right\vert }{R}\left( ~\text{\ }0\;,\;%
\frac{1}{R}(x\mathbf{\cdot }p+y\mathbf{\cdot }k)\mathbf{,\;~\;}p^{m}~~,~~~%
\mathbf{k}^{I}~\right) .  \label{AdSS2}
\end{eqnarray}
To arrive at this form two Sp$\left( 2,R\right) $ gauge choices have been
made and two Sp$\left( 2,R\right) $ constraints $X\cdot X=X\cdot P=0$ have
been solved explicitly, and the solution is parameterized by $\left( x^{m},%
\mathbf{y}^{I}\right) $. One of the Sp$\left( 2,R\right) $ gauge choices is $%
P^{+^{\prime }}\left( \tau \right) =0$ for all $\tau ,$ and the second gauge
choice is $\sqrt{X^{I}\left( \tau \right) X^{I}\left( \tau \right) }=R$ with
a $\tau $ independent $R.$ Note that $R$ is a modulus that arises from the
degrees of freedom in the larger phase space of 2T-physics. The $n+1$
coordinates $X^{I}\left( \tau \right) =R\frac{\mathbf{y}^{I}}{\left\vert
\mathbf{y}\right\vert }$ which consist of a unit vector times the constant
radius $R,$ represent motion on the sphere $S^{n}$ embedded in an $n+1$
dimensional volume. The constraint $0=X\cdot X=\left( X^{I}\right) ^{2}+%
\tilde{X}^{2}$ implies that the $\left( d-n+1\right) $ coordinates $\tilde{X}%
=\left( X^{+^{\prime }},X^{-^{\prime }},X^{m}\right) ,$ which have signature
$\left( \left( d-n-1\right) ,2\right) ,$ satisfy $\tilde{X}\cdot \tilde{X}%
=-R^{2}$. The solution of the constraint $\tilde{X}\cdot \tilde{X}=-R^{2},$
which automatically represents motion on the space AdS$_{d-n},$ is
parameterized by the $d-n$ coordinates $\left( x^{m~},\left\vert \mathbf{y}%
\right\vert \right) $ as given above.

For $d=4$ we have a total of four dimensions $\left( x^{m}\left( \tau
\right) ,y^{I}\left( \tau \right) \right) $ that remain after gauge fixing.
In this gauge the flat $(4,2)$ metric generates the AdS$_{4-n}\times $S$^{n}$
metric,

\begin{eqnarray}
ds^{2} &=&dX^{M}dX^{N}\eta _{MN}=\frac{R^{2}}{y^{2}}\left[ \left(
dx^{m}\right) ^{2}+\left( d\mathbf{y}^{I}\right) ^{2}\right] \\
&=&\frac{R^{2}}{y^{2}}\left[ \left( dx^{m}\right) ^{2}+\left( dy\right) ^{2}%
\right] +R^{2}\left( d\mathbf{\Omega }\right) ^{2},  \label{AdSSmetric}
\end{eqnarray}%
where we have decomposed $\mathbf{y}^{I}$ in spherical coordinates $\mathbf{y%
}^{I}=y\mathbf{\Omega }^{I}$ into its radial $y=\left\vert \mathbf{y}%
\right\vert $ and angular $\mathbf{\Omega }^{I}=\frac{\mathbf{y}^{I}}{%
\left\vert \mathbf{y}\right\vert }$ parts$,$ and wrote $\left( d\mathbf{y}%
\right) ^{2}=\left( dy\right) ^{2}+y^{2}\left( d\mathbf{\Omega }\right)
^{2}. $ Evidently, $\frac{R^{2}}{y^{2}}\left[ \left( dx^{m}\right)
^{2}+\left( dy\right) ^{2}\right] $ is the AdS$_{4-n}$ metric and $%
R^{2}\left( d\mathbf{\Omega }\right) ^{2}$ is the S$^{n}$ metric. The same
parametrization can be used in $d+2$ phase space for any $d$ to construct AdS%
$_{d-n}\times $S$^{n}$ \cite{2tHandAdS}\cite{2tAdSs}.

In the case of AdS$_{4}$ we take $n=0$ which corresponds to the following
coordinates

\begin{eqnarray}
X^{M} &=&\left( \frac{\overset{+^{\prime }}{R^{2}}}{y},\;~~~\overset{%
-^{\prime }}{\frac{x\cdot x+y^{2}}{2y}},~~\;~~~~\overset{m=0,1,2}{\frac{%
Rx^{m}}{y},}~~\overset{}{R}\right)  \label{AdS41} \\
P^{M} &=&\left( ~\text{\ \ }0\;,\;\frac{y}{R^{2}}(x\mathbf{\cdot }p+yk)%
\mathbf{,\;\;}\frac{p^{m}y}{R},~~~\frac{ky}{R}\right) .  \label{AdS42}
\end{eqnarray}%
where $X^{3}\left( \tau \right) =R,$ is the gauge choice and $X^{m}\left(
\tau \right) =\frac{Rx^{m}\left( \tau \right) }{y\left( \tau \right) },$ $%
m=0,1,2$. The dot products with $x^{m}\left( \tau \right) ,p^{m}\left( \tau
\right) $ involves the 3-dimensional Minkowski metric $\eta ^{mn}$. The
structure of $X^{M},P^{M}$ is such that the 2T-physics action reduces to the
action of a particle moving on AdS$_{4}$%
\begin{eqnarray}
S &=&\int d\tau ~\left( \dot{X}^{M}P^{N}-\frac{1}{2}A^{ij}X_{i}^{M}X_{j}^{N}%
\right) \eta _{MN} \\
&=&\int d\tau \left( \dot{x}^{m}p_{m}+\dot{y}k-\frac{1}{2}A^{22}\left(
p^{m}p_{m}+k^{2}\right) \frac{y^{2}}{R^{2}}\right) .  \label{AdS4action}
\end{eqnarray}%
This justifies the parametrization given in Eqs.(\ref{AdSS1},\ref{AdSS2}).

The last factor multiplying $A^{22}$ is the Laplacian of AdS$_{d}$ at the
classical level, namely $p_{\mu }p_{\nu }g^{\mu \nu }=\left(
p^{m}p_{m}+k^{2}\right) \frac{y^{2}}{R^{2}},$ where $g^{\mu \nu }$ is the
inverse of the metric in Eq.(\ref{AdSSmetric}) for $n=0$. When it is quantum
ordered into a Hermitian form $\left[ y\left( p^{m}p_{m}+k^{2}\right) y%
\right] $ that is applied on physical states $\psi ,$ namely $y\left(
\partial ^{m}\partial _{m}+\partial _{y}^{2}\right) y\psi \left(
x^{m},y\right) =0$, this form gives the correct Laplacian for a particle on
AdS$_{d}$ after a re-scaling $\psi =\left( -g\right) ^{1/4}\phi =\left(
\frac{R}{y}\right) ^{d/2}\phi $
\begin{equation}
\frac{1}{\sqrt{-g}}\partial _{\mu }\left( \sqrt{-g}g^{\mu \nu }\partial
_{\nu }\phi \right) +\frac{d\left( d-2\right) }{4R^{2}}\phi =0.
\end{equation}%
The quantum ordering introduces a quantized mass term $m_{\phi }^{2}=d\left(
d-2\right) /4R^{2}$ which is required by the SO$\left( d,2\right) $
invariance of the AdS$_{d}$ particle at the quantum level, or by the
corresponding field theory, as shown in (\cite{2tHandAdS}). Note that the SO$%
\left( d,2\right) $ symmetry (SO$\left( 4,2\right) $ in the case of AdS$_{4}$%
) is larger than the commonly mentioned SO$\left( d-1,2\right) $ symmetry of
AdS$_{d}$ space (SO$\left( 3,2\right) $ in the case of AdS$_{4}$). The extra
symmetry (which is noticeable from the left side of Eq.(\ref{AdSSmetric}))
is hidden in the $\left( x_{m},y\right) $ basis and its presence was first
noticed through the 2T-physics formulation \cite{2tHandAdS}. This symmetry
becomes evident also in the twistor basis below.

\bigskip Inserting the coordinates $X^{\mu }=\frac{R}{y}\left(
x^{m},y\right) ,$ $P^{\mu }=\frac{y}{R}\left( p^{m},k\right) $ and $%
X^{+^{\prime }}=\frac{R^{2}}{y},$ $P^{+^{\prime }}=0$ in the twistor
transform of Eq.(\ref{genpenrose}) we obtain the twistors for the AdS$_{4}$
particle,%
\begin{eqnarray}
\mu ^{\dot{\alpha}} &=&-i\frac{X^{\dot{\alpha}\beta }}{X^{+^{\prime }}}%
\lambda _{\beta }=-i\frac{y}{\sqrt{2}R^{2}}\left( \frac{R}{y}x^{m}\bar{\sigma%
}_{m}+R\sigma _{3}\right) ^{\dot{\alpha}\beta }\lambda _{\beta } \\
&=&-i\frac{1}{\sqrt{2}R}\left( x^{m}\bar{\sigma}_{m}+y\sigma _{3}\right) ^{%
\dot{\alpha}\beta }\lambda _{\beta },\;\;
\end{eqnarray}%
and
\begin{eqnarray}
\lambda _{\alpha }\bar{\lambda}_{\dot{\beta}} &=&X^{+^{\prime }}P_{\alpha
\dot{\beta}}=\frac{R^{2}}{\sqrt{2}y}\left( \frac{y}{R}p^{m}\sigma _{m}+\frac{%
y}{R}k\sigma _{3}\right) \\
&=&\frac{R}{\sqrt{2}}\left( p^{m}\sigma _{m}+k\sigma _{3}\right) .
\end{eqnarray}%
Thus, in this gauge the twistor structure is very similar to the massless
particle in $d=4$, with $x^{\mu }=\left( x^{m},y\right) $ and $p^{\mu
}=\left( p^{m},k\right) ,$ while the form of the twistor $Z=\left( \QATOP{%
\mu }{\lambda }\right) $ displays the SU$\left( 2,2\right) =$SO$\left(
4,2\right) $ symmetry in the fundamental representation.

Let's examine the canonical structure of phase space induced by the
canonical structure of the twistors, as described in Eq.(\ref{tw22})

\begin{eqnarray}
S_{0} &=&i\int d\tau \bar{Z}_{A}\partial _{\tau }Z^{A}=i\int d\tau \left[
\bar{\lambda}_{\dot{\alpha}}\partial _{\tau }\mu ^{\dot{\alpha}}+\bar{\mu}%
^{\alpha }\partial _{\tau }\lambda _{\alpha }\right] \\
&=&\int d\tau \frac{\partial }{\partial \tau }\left( \frac{X_{\mu }}{%
X^{+^{\prime }}}\right) \left( X^{+^{\prime }}P_{\mu }\right) \\
&=&\int d\tau \frac{\partial }{\partial \tau }\left( \frac{x^{\mu }}{R^{2}}%
\right) \left( Rp_{\mu }\right) =\int d\tau \left( \dot{x}^{m}p_{m}+\dot{y}%
k\right)
\end{eqnarray}%
So, the canonical structure is correct. Next we examine the on-shell
condition%
\begin{equation}
0=\left( P^{\mu }-\frac{P^{+^{\prime }}}{X^{+^{\prime }}}X^{\mu }\right)
^{2}=P^{2}=\frac{y^{2}}{R^{2}}\left( p^{m}p_{m}+k^{2}\right) =0.
\end{equation}%
This gives the correct Laplacian for the massless conformal particle in the
curved AdS$_{4}$ space as explained after Eq.(\ref{AdS4action}). The
positivity condition trivially gives $p^{0}>0.$

Next we examine the inverse relation
\begin{equation}
\left( Z\bar{Z}\right) _{~B}^{A}=\left(
\begin{array}{cc}
\mu ^{\dot{\alpha}}\bar{\lambda}_{\dot{\beta}} & \mu ^{\dot{\alpha}}\bar{\mu}%
^{\beta } \\
\lambda _{\alpha }\bar{\lambda}_{\dot{\beta}} & \lambda _{\alpha }\bar{\mu}%
^{\beta }%
\end{array}%
\right) =\frac{1}{2i}\left(
\begin{array}{cc}
{L}^{+^{\prime }-^{\prime }}{+}\frac{1}{2}{L}^{\mu \nu }{\bar{\sigma}}_{\mu
\nu } & -i\sqrt{{2}}{L}^{-^{\prime }\mu }~{\bar{\sigma}}_{\mu } \\
i\sqrt{{2}}{L}^{+^{\prime }\mu }~{\sigma }_{\mu } & {-L}^{+^{\prime
}-^{\prime }}{+}\frac{1}{2}{L^{\mu \nu }\sigma }_{\mu \nu }%
\end{array}%
\right)
\end{equation}%
It can be verified that the twistors correctly reproduce the $L^{MN}$
constructed from the vectors in Eq.(\ref{AdS41},\ref{AdS42}), as shown in
Eqs.(\ref{lmnclass1}-\ref{lmnclass5}) below, and that they correctly close
into the SO$\left( 4,2\right) $ Lie algebra at the classical level. Hence we
have constructed the correct twistor description of the AdS$_{4}$ particle.

The twistor version explicitly shows that the AdS$_{4}$ particle has the SO$%
\left( 4,2\right) $ symmetry which is larger than the commonly expected SO$%
\left( 3,2\right) $ symmetry. The larger symmetry is also evident by writing
the metric in the form of Eq.(\ref{AdSSmetric}) and noting the symmetries of
the left hand side $ds^{2}=dX^{M}dX^{N}\eta _{MN}.$ The reason for the extra
symmetry is in the fact that $R$ is a gauge fixed form of an extra
coordinate as in Eq.(\ref{AdS41}), and this is not noticed in common
approaches in discussing AdS symmetries. In any case our AdS$_{4}$ twistor
transform makes it evident that the system has the SU$\left( 2,2\right) $
symmetry linearly realized in the twistor version in Eq.(\ref{action}) or in
the 2T-physics version in Eq.(\ref{2Taction}).

We now turn to the cases of AdS$_{4-n}\times $S$^{n}$ for $n=1,2$ and
compute the canonical structure induced by the twistors in Eq.(\ref{tw22}), $%
S_{0}=\int d\tau \frac{\partial }{\partial \tau }\left( \frac{\left\vert
\mathbf{y}\right\vert X^{\mu }}{R^{2}}\right) \left( \frac{R^{2}}{\left\vert
\mathbf{y}\right\vert }P_{\mu }\right) .$ We write $x^{\mu }=\left(
x^{m},y^{I}\right) ,$ $p_{\mu }=\left( p_{m},k_{I}\right) $ with $I=1,\cdots
,n+1,$ and use the definitions $X^{\mu }=\frac{R}{\left\vert \mathbf{y}%
\right\vert }x^{\mu }$ , $P_{\mu }=p_{\mu }\frac{\left\vert \mathbf{y}%
\right\vert }{R}$ that follows from Eqs.(\ref{AdSS1},\ref{AdSS2})$. $ Then
we obtain the correct canonical structure

\begin{equation*}
S_{0}=\int d\tau \frac{\partial }{\partial \tau }\left( \frac{\left\vert
\mathbf{y}\right\vert X^{\mu }}{R^{2}}\right) \left( \frac{R^{2}}{\left\vert
\mathbf{y}\right\vert }P_{\mu }\right) =\int d\tau \frac{\partial }{\partial
\tau }\left( \frac{x^{\mu }}{R}\right) \left( Rp_{\mu }\right) =\int d\tau
\left( \dot{x}^{m}p_{m}+\mathbf{\dot{y}}^{I}\mathbf{k}^{I}\right) .
\end{equation*}%
Similarly, the mass-shell condition in Eq.(\ref{tw33}) gives%
\begin{eqnarray}
0 &=&\left( P^{\mu }-\frac{P^{+^{\prime }}}{X^{+^{\prime }}}X^{\mu }\right)
^{2}=P^{2}=\frac{y^{2}}{R^{2}}\left( p^{m}p_{m}+\mathbf{k}^{2}\right) \\
&=&\frac{y^{2}}{R^{2}}\left( p^{m}p_{m}+k^{2}\right) +\frac{1}{2}L^{IJ}L_{IJ}
\end{eqnarray}%
where in the last line we rewrote the vector dot product $\mathbf{k}^{2}$ in
terms of the radial and angular momenta. The Casimir operator $\frac{1}{2}%
L^{IJ}L_{IJ}$ for SO$\left( n+1\right) $ is the correct Laplacian on S$^{n}.$
After quantum ordering into a Hermitian operator $y\left(
p^{m}p_{m}+k^{2}\right) y+\frac{1}{2}L^{IJ}L_{IJ}=0$ applied on physical
states $\psi \left( x^{m},y,\mathbf{\Omega }\right) =\left( -g\right)
^{1/4}\phi $, this gives the correct Laplacian for the massless particle in
the curved AdS$_{d-n}\times $S$^{n}$ space for the metric of Eq.(\ref%
{AdSSmetric}), with an induced mass $m_{\phi }^{2}=\left( d-2n\right) \left(
d-2\right) /4R^{2}$ at the quantum level, as discussed in \cite{2tHandAdS}.
Note that the mass vanishes for the case of $d=2n,$ which applies to AdS$%
_{2}\times $S$^{2}$ in the present case. The positivity condition $\left(
X^{+^{\prime }}P^{0}-P^{+^{\prime }}X^{0}\right) >0$ gives trivially $%
p^{0}>0.$

Similarly to the AdS$_{4}$ case above the inverse relations give the SO$%
\left( 4,2\right) $ generators consistently in terms of twistors or in terms
of the gauge fixed vectors $X^{M},P^{M}.$ These are given by for every $%
n=0,1,2$ as
\begin{eqnarray}
L^{IJ} &=&\mathbf{y}^{I}\mathbf{k}^{J}-\mathbf{y}^{J}\mathbf{k}^{I},\quad
L^{\mu \nu }=x^{\mu }p^{\nu }-x^{\nu }p^{\mu },  \label{lmnclass1} \\
L^{+^{\prime }-^{\prime }} &=&\mathbf{y\cdot k}+x\cdot p,\quad L^{+^{\prime
}\mu }=p^{\mu },\quad L^{+^{\prime }I}=\mathbf{k}^{I},\quad \\
L^{-^{\prime }\mu } &=&\frac{1}{2}\left( x^{2}+\mathbf{y}^{2}\right) p^{\mu
}-\left( x\cdot p+\mathbf{y\cdot k}\right) x^{\mu },  \label{lmnclass} \\
L^{-^{\prime }I} &=&\frac{1}{2}\left( x^{2}+\mathbf{y}^{2}\right) \mathbf{k}%
^{I}-\left( x\cdot p+\mathbf{y\cdot k}\right) \mathbf{y}^{I}, \\
L^{\mu I} &=&x^{\mu }\mathbf{\,k}^{I}-p^{\mu }\mathbf{y}^{I}.
\label{lmnclass5}
\end{eqnarray}%
Thus the spaces AdS$_{4-n}\times $S$^{n}$ have hidden symmetries SO$\left(
4,2\right) $ which is larger than the commonly discussed SO$\left(
3-n,2\right) \times $SO$\left( n+1\right) .$ Again the crucial observation
in understanding what is missed in common discussions, is that $R$ is a
gauge fixed form of an additional coordinate that corresponds to the AdS$%
\times $S radius $\sqrt{X^{I}\left( \tau \right) X^{I}\left( \tau \right) }%
=R $. The full SO$\left( 4,2\right) $ acts non-linearly on the remaining
degrees of freedom after the gauge is fixed, but it acts linearly before the
gauge fixing. The twistor transform makes the hidden symmetry evident in the
twistor version of the system. The larger hidden symmetry in phase space
version of AdS$_{d-n}\times $S$^{n}$ is discussed in detail at the quantum
level for any $d$ in the second paper\footnote{%
To adapt the quantum results of \cite{2tHandAdS} we must use a translation
of notations. Instead of the canonical variables $\left( \mathbf{y,k}\right)
$ that we used here, ref.\cite{2tHandAdS} uses the canonical set $(\mathbf{u,%
\tilde{k})}$ (but $\mathbf{\tilde{k}}$ is called $\mathbf{k}$ in \cite%
{2tHandAdS}). The relation between these parameterizations at the
quantum level is given by $\mathbf{y=u/u}^{2}$ and
$\mathbf{k=\tilde{k}u}^{2}-\left( \mathbf{\tilde{k}\cdot u+u\cdot
\tilde{k}}\right) \mathbf{u}$ as described in Eqs.(74--79) and
footnote 4 in ref.\cite{2tHandAdS}.} in \cite{2tHandAdS}.

\section{Twistors for particle on $R\times $S$^{3}$}

We make two Sp$\left( 2,R\right) $ gauge choices and solve two constraints
to obtain the following gauge fixed form of the $\left( 4+2\right) $
dimensional phase space in 2T-physics
\begin{eqnarray}
X^{M} &=&R\left( \overset{0^{\prime }}{\cos t}\;,\;\;\overset{0}{\sin t},~~~%
\overset{I=1,2,3,4}{\mathbf{\hat{r}}^{I}}\right) ,  \label{RS1} \\
P^{M} &=&\frac{1}{R}\left( -H\sin t,\;H\cos t~,\;\mathbf{\hat{r}}_{J}\mathbf{%
L}^{JI}\right) ,  \label{RS2}
\end{eqnarray}%
where $\mathbf{\hat{r}}$ is the unit vector $\mathbf{\hat{r}}\left(
\tau \right) \mathbf{=r}^{I}\left( \tau \right) /\left\vert
\mathbf{r}\right\vert $ that defines the motion on the sphere
$S^{3}$ embedded in four Euclidean dimensions, and $\left(
X^{I}X^{I}\right) ^{1/2}=\left\vert \mathbf{r}\left( \tau \right)
\right\vert =R$ is the radial coordinate in spherical coordinates
that has been gauge fixed to be a constant for all $\tau $.
Evidently $\mathbf{\hat{r}}^{I}$ can be parameterized in terms of
three angles, but we will not need to give an explicit
parametrization.

Defining the SO$\left( 4\right) $ rotation generators $L^{IJ}=\mathbf{r}^{I}%
\mathbf{p}^{J}-$ $\mathbf{r}^{J}\mathbf{p}^{I}$, we note that $\frac{1}{R}%
\mathbf{\hat{r}}_{J}\mathbf{L}^{JI}=\mathbf{p}^{I}-\mathbf{\hat{r}\hat{r}%
\cdot p}$ is purely SO$\left( 4\right) $ angular momentum since the radial
momentum $\mathbf{\hat{r}\cdot p}$ has been subtracted. In fact $\mathbf{%
\hat{r}\cdot p}$ drops out everywhere, which is equivalent to choosing the
second gauge as $\mathbf{\hat{r}\cdot p=}0\mathbf{\ }$for all $\tau .$ Then
the Sp$\left( 2,R\right) $ constraints $X^{2}=X\cdot P=0$ are explicitly
solved with the above form. In this gauge the 2T action in Eq.(\ref{2Taction}%
) reduces to%
\begin{eqnarray}
S &=&\int d\tau ~\left( \dot{X}^{M}P^{N}-\frac{1}{2}A^{ij}X_{i}^{M}X_{j}^{N}%
\right) \eta _{MN} \\
&=&\int d\tau \left( -H\partial _{\tau }t+\mathbf{L}^{JI}\mathbf{\hat{r}}%
_{J}\partial _{\tau }\mathbf{\hat{r}}_{I}-\frac{1}{2R^{2}}A^{22}\left(
-H^{2}+\frac{1}{2}L^{IJ}L_{IJ}\right) \right)  \label{rs3}
\end{eqnarray}%
where we have used $\left( \mathbf{\hat{r}}_{J}\mathbf{L}^{JI}\right) ^{2}=%
\frac{1}{2}L^{IJ}L_{IJ}.$ The first two terms define the canonical structure
that gives the quantum rules for the particle on $R\times S^{3}$, namely $%
\left[ t,H\right] =-i$, $\left[ \mathbf{L}_{IJ},\mathbf{\hat{r}}_{K}\right]
=i\mathbf{\hat{r}}_{J}\delta _{IK}-i\mathbf{\hat{r}}_{I}\delta _{JK}$ and $%
\left[ \mathbf{L}_{IJ},\mathbf{L}_{KL}\right] =$SO$\left( 4\right) $ Lie
algebra. The remaining constraint
\begin{equation}
P\cdot P=\frac{1}{R^{2}}\left( -H^{2}+\left( \mathbf{\hat{r}}_{J}\mathbf{L}%
^{JI}\right) ^{2}\right)
\end{equation}%
appears as the coefficient of the gauge field $A^{22}.$ We could fix the
remaining gauge symmetry by fixing the gauge $t\left( \tau \right) =\tau ,$
and solving the constraint explicitly $H^{2}=\frac{1}{2}L^{IJ}L_{IJ}.$ But
we will work more generally without choosing this gauge and impose the
constraint on physical states. Then the physical states are the symmetric
traceless SO$\left( 4\right) $ tensors $T_{I_{1}\cdots I_{l}}\left( \mathbf{%
\hat{r}}\right) ,$ for which $H^{2}$ takes the values of the Casimir
operator $H^{2}=\frac{1}{2}L^{IJ}L_{IJ}=l\left( l+2\right) ,$ $%
l=0,1,2,\cdots $ of the SO$\left( 4\right) $ rotations on $S^{3}.$ So, in
this gauge the 2T-physics system is interpreted as a particle moving on $%
R\times $S$^{3}.$

We now construct the twistors for this system by applying the general
formulas in Eq.(\ref{genpenrose})%
\begin{equation}
\mu ^{\dot{\alpha}}=-i\frac{X^{\dot{\alpha}\beta }}{X^{+^{\prime }}}\lambda
_{\beta }\,,\;\;\lambda _{\alpha }\bar{\lambda}_{\dot{\beta}}=\left(
X^{+^{\prime }}P_{\alpha \dot{\beta}}-P^{+^{\prime }}X_{\alpha \dot{\beta}%
}\right)
\end{equation}%
where as usual $X^{\dot{\alpha}\beta }=\frac{1}{\sqrt{2}}X^{\mu }\left( \bar{%
\sigma}_{\mu }\right) ^{\dot{\alpha}\beta }$ and $P_{\alpha \dot{\beta}%
}\equiv \frac{1}{\sqrt{2}}P^{\mu }\left( \sigma _{\mu }\right) _{\alpha \dot{%
\beta}}$, and following the definitions in Eqs.(\ref{RS1},\ref{RS2}) we
identify $X^{+^{\prime }},P^{+^{\prime }},X^{\mu },P^{\mu }$ as follows
\begin{eqnarray}
X^{+^{\prime }} &=&\frac{R}{\sqrt{2}}\left( \cos t+\mathbf{\hat{r}}%
_{4}\right) ,\;\;\;X^{\mu }=R\left( \overset{0}{\sin t}~~\overset{i=1,2,3}{%
\mathbf{\hat{r}}^{i}}\right) ,\;\left( \mathbf{\hat{r}}_{i}\right)
^{2}+\left( \mathbf{\hat{r}}_{4}\right) ^{2}=1 \\
\;P^{+^{\prime }} &=&\frac{1}{\sqrt{2}R}\left( -H\sin t+\mathbf{\hat{r}}^{j}%
\mathbf{L}_{j4}\right) ,\;P^{\mu }=\frac{1}{R}\left( \overset{0}{H\cos t}~~%
\overset{i=1,2,3}{\mathbf{\hat{r}}^{j}\mathbf{L}_{ji}}\right)
\end{eqnarray}%
We could parameterize $\mathbf{\hat{r}}_{4}=\cos \theta ,$ $\mathbf{\hat{r}}%
^{i}=n^{i}\sin \theta ,$ where $n^{i}$ is a unit vector in 3 dimensions, but
we will not need such an explicit parametrization. With the above
definitions we find the twistor transform for the particle on $S^{3}$ as
follows%
\begin{eqnarray}
\mu ^{\dot{\alpha}} &=&-i\frac{\sqrt{2}\left( -\sin t+\mathbf{\hat{r}}%
^{i}\sigma ^{i}\right) ^{\dot{\alpha}\beta }}{\left( \cos t+\mathbf{\hat{r}}%
_{4}\right) }\lambda _{\beta },\;\; \\
\lambda _{\alpha }\bar{\lambda}_{\dot{\beta}} &=&\frac{1}{\sqrt{2}}\left[
\begin{array}{c}
\left( \cos t+\mathbf{\hat{r}}_{4}\right) \left( H\cos t+\mathbf{\hat{r}}^{j}%
\mathbf{L}_{ji}\sigma ^{i}\right) \\
-\left( -H\sin t+\mathbf{\hat{r}}^{j}\mathbf{L}_{j4}\right) \left( \sin t+%
\mathbf{\hat{r}}^{i}\sigma ^{i}\right)%
\end{array}%
\right] _{\alpha \dot{\beta}}
\end{eqnarray}%
Next we check the canonical structure%
\begin{eqnarray}
S_{0} &=&i\int d\tau \bar{Z}_{A}\partial _{\tau }Z^{A}=i\int d\tau \left[
\bar{\lambda}_{\dot{\alpha}}\partial _{\tau }\mu ^{\dot{\alpha}}+\bar{\mu}%
^{\alpha }\partial _{\tau }\lambda _{\alpha }\right] \\
&=&\int d\tau Tr\left\{ \partial _{\tau }\left( \frac{X^{\mu }}{X^{+^{\prime
}}}\right) \left( X^{+^{\prime }}P_{\mu }-P^{+^{\prime }}X_{\mu }\right)
\right\} \\
&=&\int d\tau \left( -H\partial _{\tau }t+\mathbf{L}^{JI}\mathbf{\hat{r}}%
_{J}\partial _{\tau }\mathbf{\hat{r}}_{I}\right)
\end{eqnarray}%
This is the correct canonical structure for $R\times S^{3}$, as described
following Eq.(\ref{rs3}).

Turning to the on-shell condition we find $\det \left( \lambda \bar{\lambda}%
\right) =0$ implies%
\begin{eqnarray}
0 &=&\left( P_{\mu }-\frac{P^{+^{\prime }}}{X^{+^{\prime }}}X_{\mu }\right)
^{2}  \notag \\
&=&-\frac{1}{R^{2}}\left( H\cos t-\frac{-H\sin t+\mathbf{\hat{r}}^{j}\mathbf{%
L}_{j4}}{\cos t+\mathbf{\hat{r}}_{4}}\sin t\right) ^{2}+\frac{1}{R^{2}}%
\left( \mathbf{\hat{r}}^{j}\mathbf{L}_{ji}-\frac{-H\sin t+\mathbf{\hat{r}}%
^{j}\mathbf{L}_{j4}}{\cos t+\mathbf{\hat{r}}_{4}}\mathbf{\hat{r}}^{i}\right)
^{2}  \notag \\
&=&\frac{1}{R^{2}}\left( -H^{2}+\frac{1}{2}L^{IJ}L_{IJ}\right) ,
\end{eqnarray}%
which imposes the correct constraint $H^{2}=\frac{1}{2}L^{IJ}L_{IJ}.$

Finally, the sign condition is%
\begin{eqnarray}
\left[ \left( \cos t+\mathbf{\hat{r}}_{4}\right) H\cos t+\left( H\sin t-%
\mathbf{\hat{r}}^{j}\mathbf{L}_{j4}\right) \sin t\right] &>&0 \\
\text{ or }\left[ H+H\mathbf{\hat{r}}_{4}\cos t-\mathbf{\hat{r}}^{j}\mathbf{L%
}_{j4}\sin t\right] &>&0
\end{eqnarray}%
This is satisfied for all $t.$ Therefore the twistors correctly describe the
particle on $R\times $S$^{3}.$

\section{Twistors for H-atom\label{other}}

For the H-atom gauge let us consider the following gauge choice \cite%
{2tHandAdS} for the 4+2 dimensional phase space

\begin{eqnarray}
X^{M} &=&F\left( \overset{0^{\prime }}{\cos u~},\;\overset{1^{\prime }}{-%
\frac{1}{\alpha }\sqrt{-2H}\mathbf{r\cdot p}},~\overset{0}{\sin u~},~~%
\overset{i}{(\frac{1}{r}\mathbf{r}^{i}-\frac{\mathbf{r\cdot p}}{\alpha }%
\mathbf{p}^{i})~}\right)  \label{H-atom1} \\
P^{M} &=&G\left( -\sin u,\;\;\;\;~(1-\frac{r\mathbf{p}^{2}}{\alpha })\mathbf{%
\;\;,\;\cos }u\mathbf{~,\;~~~\;}\sqrt{-2H}\frac{r}{\alpha }\mathbf{\mathbf{p}%
}^{i}~\mathbf{~~~}\right)  \label{H-atom2} \\
GF &=&\frac{\alpha }{\sqrt{-2H}},~~~~~u=\frac{\sqrt{-2H}}{\alpha }\left(
\mathbf{r\cdot p}-2\tau H\right) ,\;~H=\frac{\mathbf{p}^{2}}{2}-\frac{\alpha
}{r},
\end{eqnarray}%
where we have fixed the three gauge degrees of freedom and imposed all three
constraints $X^{2}=P^{2}=X\cdot P=0$. Using these coordinates the 2T action
in Eq.(\ref{2Taction}) reduces to%
\begin{eqnarray}
S &=&\int d\tau ~\left( \dot{X}^{M}P^{N}-\frac{1}{2}A^{ij}X_{i}^{M}X_{j}^{N}%
\right) \eta _{MN} \\
&=&\int d\tau \left( \mathbf{p}^{i}\partial _{\tau }\mathbf{r}_{i}-H\right) .
\end{eqnarray}%
This is the non-relativistic H-atom action in three space dimensions. Now
that we have introduced the 2T-physics gauge, we are going to consider the
twistor version for this case. As in the previous cases, we construct the
twistors for this system by applying the general formulas in Eq.(\ref%
{genpenrose})%
\begin{equation}
\mu ^{\dot{\alpha}}=-i\frac{X^{\dot{\alpha}\beta }}{X^{+^{\prime }}}\lambda
_{\beta }\,,\;\;\lambda _{\alpha }\bar{\lambda}_{\dot{\beta}}=\left(
X^{+^{\prime }}P_{\alpha \dot{\beta}}-P^{+^{\prime }}X_{\alpha \dot{\beta}%
}\right)
\end{equation}%
where the coordinates $X^{+^{\prime }},P^{+^{\prime }},~X^{\mu },P^{\mu }$
come from (\ref{H-atom1},\ref{H-atom2}) as
\begin{eqnarray}
X^{+^{\prime }} &=&\frac{F}{\sqrt{2}}\left( \cos u-\frac{1}{\alpha }\sqrt{-2H%
}\mathbf{r\cdot p}\right) ,\;~X^{\mu }=F\left( \overset{0}{\sin u}~~\overset{%
i=1,2,3}{\frac{1}{r}\mathbf{r}^{i}-\frac{\mathbf{r\cdot p}}{\alpha }\mathbf{p%
}^{i}}\right) ,\; \\
\;P^{+^{\prime }} &=&\frac{G}{\sqrt{2}}\left( -\sin u+~1-\frac{r\mathbf{p}%
^{2}}{\alpha }\right) ,\;~~~~P^{\mu }=G\left( \overset{0}{\cos u}~~\overset{%
i=1,2,3}{\sqrt{-2H}\frac{r}{\alpha }\mathbf{\mathbf{p}}^{i}}\right) .
\end{eqnarray}%
So the twistor transform becomes%
\begin{eqnarray}
\mu ^{\dot{\alpha}} &=&-i\frac{\left( -\sin u+\frac{1}{r}\mathbf{r}^{i}%
\mathbf{\sigma }^{i}-\frac{\mathbf{r\cdot p}}{\alpha }\mathbf{p}^{i}\mathbf{%
\sigma }^{i}\right) ^{\dot{\alpha}\beta }}{\cos u-\frac{1}{\alpha }\sqrt{-2H}%
\mathbf{r\cdot p}}\lambda _{\beta } \\
\lambda _{\alpha }\bar{\lambda}_{\dot{\beta}} &=&\frac{\alpha }{2\sqrt{-2H}}%
\left(
\begin{array}{c}
\left( \cos u-\frac{1}{\alpha }\sqrt{-2H}\mathbf{r\cdot p}\right) \left(
\cos u+\sqrt{-2H}\frac{r}{\alpha }\mathbf{\mathbf{p}}^{i}\mathbf{\sigma }%
^{i}\right) \\
-\left( -\sin u+~1-\frac{r\mathbf{p}^{2}}{\alpha }\right) \left( \sin u+%
\frac{1}{r}\mathbf{r}^{i}\mathbf{\sigma }^{i}-\frac{\mathbf{r\cdot p}}{%
\alpha }\mathbf{p}^{i}\mathbf{\sigma }^{i}\right)%
\end{array}%
\right) _{\alpha \dot{\beta}}
\end{eqnarray}%
If we substitute the twistor in this case in Eq. (\ref{tw22}), we can see
that the canonical structure reduces to the Lagrangian for the Hydrogen
atom, plus a total derivative, $\partial _{\tau }\left( 3\tau H-2\mathbf{%
r\cdot p}\right) $, that is dropped in the last line below%
\begin{eqnarray}
S_{0} &=&i\int d\tau \bar{Z}_{A}\partial _{\tau }Z^{A}=i\int d\tau \left[
\bar{\lambda}_{\dot{\alpha}}\partial _{\tau }\mu ^{\dot{\alpha}}+\bar{\mu}%
^{\alpha }\partial _{\tau }\lambda _{\alpha }\right] \\
&=&\int d\tau \frac{\partial }{\partial \tau }\left( \frac{X^{\mu }}{%
X^{+^{\prime }}}\right) \left( X^{+^{\prime }}P_{\mu }-P^{+^{\prime }}X_{\mu
}\right) \\
&=&\int d\tau \left[ \partial _{\tau }X^{\mu }P_{\mu }-\frac{P^{+^{\prime }}%
}{X^{+^{\prime }}}\partial _{\tau }X^{\mu }X_{\mu }+\left( X^{+^{\prime
}}X^{\mu }P_{\mu }-P^{+^{\prime }}X^{\mu }X_{\mu }\right) \partial _{\tau }%
\frac{1}{X^{+^{\prime }}}\right] \\
&=&\int d\tau \left( \mathbf{\dot{r}\cdot \mathbf{p-}}H\mathbf{\mathbf{+}}%
\partial _{\tau }\left( 3\tau H-2\mathbf{r\cdot p}\right) \right) \\
&\rightarrow &\int d\tau \left( \mathbf{\dot{r}\cdot \mathbf{p-}}H\right)
\end{eqnarray}%
Also, the mass shell condition $\left( P_{\mu }-X_{\mu }P^{+^{\prime
}}/X^{+^{\prime }}\right) ^{2}=0$ that is required by $\det \left( \lambda
\bar{\lambda}\right) =0$ is fulfilled for the twistors above when $H=\frac{%
\mathbf{p}^{2}}{2}-\frac{\alpha }{r}$.

Finally, the positivity condition $X^{+^{\prime }}P_{0}-P^{+^{\prime
}}X_{0}>0$ takes the form%
\begin{equation}
\left[ 1-\frac{\sqrt{-2H}}{\alpha }\mathbf{r\cdot p}\cos u\mathbf{~-}\left(
~1-\frac{r\mathbf{p}^{2}}{\alpha }\right) \sin u\right] >0.
\end{equation}%
This inequality can be written in terms of the dimensionless $v\equiv \frac{%
rp^{2}}{\alpha }$ with $v\leq 2,$ and and the angle $\mathbf{\hat{r}\cdot
\hat{p}=}\cos \theta $ as follows
\begin{equation}
\left[ 1-\sqrt{2v-v^{2}}\cos \theta \cos u-\left( 1-v\right) \sin u\right] >0
\end{equation}%
This is satisfied for any $u\left( \tau \right) ,$ and therefore for any $%
\tau .$ Hence the twistors given above correctly describe the H-atom.

\section{Conclusions and comments}

We have constructed the twistors for an assortment of particle dynamical
systems, including special examples of massless or massive particles,
relativistic or non-relativistic, interacting or non-interacting, in flat
space or curved spaces. More examples can be constructed in one to one
correspondence with all other possible gauge choices that we can make in
2T-physics. Our unified construction involves always the \textit{same}
twistor $Z^{A}$ with only four complex degrees of freedom and subject to the
\textit{same} helicity constraint. Only the twistor to phase space transform
differs from one case to another. Hence a unification of diverse particle
dynamical systems is displayed by the fact that they all share the same
twistor description.

Of course, this unification is equivalent to 2T-physics, except that in the
present case it is expressed in terms of twistors instead of the 6
dimensional vectors $X^{M},P^{M}.$ Furthermore, the actions in the six
dimensional phase space Eq.(\ref{2Taction}) or in the twistor space Eq.(\ref%
{action}) are both SO$\left( 4,2\right) =$SU$\left( 2,2\right) $ invariant
and are physically equivalent. Either form of the action can be taken as the
starting point to derive all of the results of this paper. The equivalence
of the two actions is derived as two gauge fixed forms of the same
2T-physics action, one in the \textquotedblleft particle
gauge\textquotedblright , and the other in the \textquotedblleft twistor
gauge\textquotedblright , as explained in \cite{2ttwistor}\cite%
{2tsuperstring}\cite{2tstringtwistors}\cite{twistorD}.

The twistor to phase space transform for the cases of $R\times $S$^{3}$ and
H-atom seemed rather complicated. One of the reasons for this is that the
natural evident symmetry for these cases is SO$\left( 4\right) ,$ but the
twistor components $Z^{A}=\left( \QATOP{\mu ^{\dot{\alpha}}}{\lambda
_{\alpha }}\right) $ are expressed in an SL$\left( 2,C\right) =$SO$\left(
3,1\right) $ basis. The clash of the SO$\left( 4\right) $ versus SO$\left(
3,1\right) $ spinor bases makes the expressions for the twistor transform
complicated. It is certainly possible to choose an SO$\left( 4\right) =$SU$%
\left( 2\right) \times $SU$\left( 2\right) $ basis to express the twistor
components. This SU$\left( 2\right) \times $SU$\left( 2\right) $ is the
natural compact subgroup of SU$\left( 2,2\right) $ in the fundamental basis
for which the metric takes the form of $\tau _{3}\times 1$ instead of the $%
\tau _{1}\times 1$ used in the SO$\left( 3,1\right) $ basis (see footnote (%
\ref{gamma})). The twistor can then be expressed as $Z^{A}=\left( \QATOP{%
a_{i}}{\bar{b}^{I}}\right) $ and $\bar{Z}_{A}=\left( \bar{a}%
^{i}~-b_{I}\right) $ where $i=1,2$ and $I=1,2$ refer to the doublets of the
two different SU$\left( 2\right) $'s. An overbar such as $\bar{b}^{I}$
implies hermitian conjugate of $b_{I}.$ Inserting these into the kinetic
term $S_{0}$ we learn from $\bar{Z}_{A}i\partial _{\tau }Z^{A}=i\bar{a}%
^{i}\partial _{\tau }a_{i}-ib_{I}\partial _{\tau }\bar{b}^{I}$ that the
canonical structure is that of positive norm harmonic oscillators%
\begin{equation}
\left[ a_{i},\bar{a}^{j}\right] =\delta _{i}^{~j},\;\;\left[ b_{I},\bar{b}%
^{J}\right] =\delta _{I}^{~J}.
\end{equation}%
Hence, twistor space corresponds to the unitary Fock space constructed from
the oscillators. Equivalently one may use coherent states that diagonalize
the oscillators$.$ The twistor to phase space transforms in the SO$\left(
4\right) $ basis can be given generally as%
\begin{eqnarray}
a_{i} &=&-i\left( \frac{X^{m}\left( \bar{\sigma}_{m}\right) _{iJ}}{%
X^{0^{\prime }}+iX^{0}}\right) \bar{b}^{J},\;~~\sigma ^{m}=(\vec{\sigma}%
,i),\;\bar{\sigma}^{m}=(\vec{\sigma},-i)=\left( \sigma ^{m}\right) ^{\dagger
},~~~~~~ \\
\;a_{i}b_{J} &=&\left( \bar{\sigma}^{m}\right) _{iJ}\left[ \left(
X_{0^{\prime }}+iX_{0}\right) P_{m}-\left( P_{0^{\prime }}+iP_{0}\right)
X_{m}\right] ,\;m=1,2,3,4~\text{for SO}\left( 4\right) .  \notag
\end{eqnarray}%
This should be compared to the SL$\left( 2,C\right) $ basis in Eq.(\ref%
{genpenrose}). Note that the factor $\frac{X^{m}\bar{\sigma}_{m}}{%
X^{0^{\prime }}+iX^{0}}$ is a unitary matrix since $X^{M}X_{M}=-\left(
X_{0^{\prime }}\right) ^{2}-\left( X_{0}\right) ^{2}+\left( X_{m}\right)
^{2}=0.$ When applied to the cases of $R\times $S$^{3}$ and H-atom the
expressions for the twistor transform are considerably simpler and more
natural (we do not give the details, but this is straightforward). So it
would be more desirable to use the SO$\left( 4\right) $ basis if one
attempts to use twistors for these or similar cases.

In this paper we concentrated on spinless particles in 4 dimensions. In
future papers we will provide a similar construction of twistors for the
corresponding spinning systems \cite{twistorsSpin} and higher dimensions
\cite{twistorDspin}.

\bigskip {\Large Acknowledgments}\textbf{\bigskip }

I. Bars was supported by the US Department of Energy under grant No.
DE-FG03-84ER40168; M. Pic\'{o}n was supported by the Spanish Ministerio de
Educaci\'{o}n y Ciencia through the grant FIS2005-02761 and EU FEDER funds,
the Generalitat Valenciana and by the EU network MRTN-CT-2004-005104
\textquotedblleft Constituents, Fundamental Forces and Symmetries of the
Universe\textquotedblright . M. Pic\'{o}n wishes to thank the Spanish
Ministerio de Educaci\'{o}n y Ciencia for his FPU research grant, and the
USC Department of Physics and Astronomy for kind hospitality.


\section{Appendix (second version of massive particle)}

A massive particle gauge is obtained by setting $P^{0^{\prime }}\left( \tau
\right) =0$ and $P^{1^{\prime }}\left( \tau \right) =m$ for all $\tau .$
Note that the mass $m$ is identified as the component of momentum in a
higher dimension. By solving explicitly the two constraints $X\cdot X=0$ and
$X\cdot P=0,$ the gauge fixed form of $X^{M},P^{M}$ become \cite{2tHandAdS}

\begin{eqnarray}
X^{M} &=&\left( -\overset{0^{\prime }}{\frac{x\cdot p}{m}a},\;\overset{%
1^{\prime }}{-\frac{x\cdot p}{m}},~~\;\overset{\mu }{x^{\mu }}\right)
,\;\;a^{2}=1+\frac{m^{2}x^{2}}{\left( x\cdot p\right) ^{2}},
\label{massive1} \\
P^{M} &=&\left( ~\;\text{\ \ \ \ \ \ }0\;\;,\;\;\;\;~m\mathbf{%
\;\;\;\;,\;\;\;\;}p^{\mu }\right) .  \label{massive2}
\end{eqnarray}%
The third constraint reduces to the mass shell condition for the massive
particle $0=P\cdot P=p^{2}+m^{2}.$ In this gauge the SO$\left( 4,2\right) $
generators $L^{MN}=X^{[M}P^{N]}$ take the form%
\begin{eqnarray}
L^{\mu \nu } &=&x^{\mu }p^{\nu }-x^{\nu }p^{\mu },\;\;\;\;L^{0^{\prime
}1^{\prime }}=-\left( x\cdot p\right) a,\; \\
L^{0^{\prime }\mu } &=&-p^{\mu }\frac{x\cdot p}{m}a,\;\text{\ }%
\;L^{1^{\prime }\mu }=-\frac{x\cdot p}{m}p^{\mu }-mx^{\mu }
\end{eqnarray}%
Next we obtain the twistor relations for the massive particle by inserting $%
X^{+^{\prime }},P^{+^{\prime }},X^{\mu },P^{\mu }$ that follow from Eqs.(\ref%
{massive1},\ref{massive2})%
\begin{equation}
X^{+^{\prime }}=-\frac{x\cdot p}{\sqrt{2}m}\left( 1+a\right)
,\;\;P^{+^{\prime }}=\frac{m}{\sqrt{2}},\;\;X^{\mu }=x^{\mu },\ P^{\mu
}=p^{\mu }  \label{PX}
\end{equation}%
into Eq.(\ref{genpenrose}). This gives
\begin{eqnarray}
\mu ^{\dot{\alpha}} &=&ix^{\dot{\alpha}\beta }\lambda _{\beta }~\frac{\sqrt{2%
}m}{x\cdot p}\left( 1+a\right) ^{-1}\,,\;\; \\
\lambda _{\alpha }\bar{\lambda}_{\dot{\beta}} &=&-p_{\alpha \dot{\beta}}%
\frac{x\cdot p}{\sqrt{2}m}\left( 1+a\right) -\frac{m}{\sqrt{2}}x_{\alpha
\dot{\beta}},
\end{eqnarray}%
We note the parallels as well as the differences compared to the massless
case in Eq.(\ref{penrose}). The zero mass limit does not seem to be smooth
for $X^{M},P^{M}$, but this is not a problem since this is only up to a Sp$%
\left( 2,R\right) $ gauge transformations. In the second massive particle
gauge discussed in section (\ref{massive}) the zero mass limit for $%
X^{M},P^{M}$ is smooth.

We know from Eq.(\ref{tw11}) that the $\bar{Z}Z=0$ constraint is already
satisfied. We now turn to the canonical structure as formulated in Eq.(\ref%
{tw22}) and compute it in the present gauge%
\begin{eqnarray}
S_{0} &=&\int d\tau \frac{\partial }{\partial \tau }\left( \frac{x^{\mu }}{%
X^{+^{\prime }}}\right) \left( X^{+^{\prime }}p_{\mu }-P^{+^{\prime }}x_{\mu
}\right) \\
&=&\int d\tau \left( \dot{x}\cdot p-\partial _{\tau }\left( x\cdot p\right)
\right)
\end{eqnarray}%
A little algebra shows that the extra total derivative term emerges when $%
X^{+^{\prime }},P^{+^{\prime }}$ are inserted in the form
\begin{equation}
P^{+^{\prime }}x\cdot \partial _{\tau }x+\left( X^{+^{\prime }}\left( x\cdot
p\right) -P^{+^{\prime }}x^{2}\right) \frac{\partial _{\tau }X^{+^{\prime }}%
}{\left( X^{+^{\prime }}\right) ^{2}}=\partial _{\tau }\left( x\cdot
p\right) .
\end{equation}%
The total derivative can be dropped, so $S_{0}=\int d\tau \left( \dot{x}%
\cdot p\right) ,$ indeed gives the correct canonical structure.

Finally we check the mass shell condition as formulated in Eq.(\ref{tw33}).
This requires%
\begin{equation}
\left( p^{\mu }-\frac{P^{+^{\prime }}}{X^{+^{\prime }}}x^{\mu }\right)
^{2}=0,\;\left( X^{+^{\prime }}p^{0}-P^{+^{\prime }}x^{0}\right) >0
\end{equation}%
Inserting $X^{+^{\prime }},P^{+^{\prime }}$ we compute%
\begin{equation}
0=\left( p^{\mu }-\frac{P^{+^{\prime }}}{X^{+^{\prime }}}x^{\mu }\right)
^{2}=\frac{1}{2}\left( p^{2}+m^{2}\right)
\end{equation}%
This is the massive particle mass shell condition.

Examining the positivity condition in Eq.(\ref{tw33}), it becomes%
\begin{equation}
m^{2}x^{0}<-\left( 1+a\right) \left( x\cdot p\right) p^{0}
\end{equation}%
and using the mass shell condition that implies $p^{0}=E=\pm \sqrt{\mathbf{p}%
^{2}+m^{2}},$ we find%
\begin{equation*}
m^{2}x^{0}<-E\left( -x^{0}E+\mathbf{x\cdot p}+sign\left( a\right) \sqrt{%
\left( x^{0}E-\mathbf{x\cdot p}\right) ^{2}+m^{2}\left( -x_{0}^{2}+\mathbf{x}%
^{2}\right) }\right) .
\end{equation*}%
An analysis of this equation shows that the equality sign can be satisfied
only for complex values of $x^{0}=\frac{E}{\mathbf{p}^{2}}\left( \mathbf{%
x\cdot p\pm }i\sqrt{\mathbf{x}^{2}\mathbf{p}^{2}-\left( \mathbf{x\cdot p}%
\right) ^{2}}\right) ,$ while the inequality is satisfied for all values of $%
x^{0},$ for either sign of $E=\pm \sqrt{\mathbf{p}^{2}+m^{2}},$ as well as
for either sign of $a.$ Hence, the twistors defined above correctly describe
the massive particle.

\end{document}